\input{psfig.sty}
\documentclass[oldversion,referee]{aa}
\usepackage{maple2e}

\newcommand{\be}{\begin{equation}}
\newcommand{\ee}{\end{equation}}
\newcommand{\bea}{\begin{eqnarray}}
\newcommand{\eea}{\end{eqnarray}}
\def\kms {\ km s$^{-1}$}
\def\msol{\ifmmode {\>M_\odot}\else {$M_\odot$}\fi}
\def\cmsq{\ifmmode {\>{\rm\ cm}^2}\else {cm$^2$}\fi}
\def\psqcm{\ifmmode {\>{\rm cm}^{-2}}\else {cm$^{-2}$}\fi}
\def\psqpc{\ifmmode {\>{\rm pc}^{-2}}\else {pc$^{-2}$}\fi}
\def\pcsq{\ifmmode {\>{\rm\ pc}^2}\else {pc$^2$}\fi}
\def\Tkev{\ifmmode{T_{\rm kev}}\else {$T_{\rm keV}$}\fi}
\def\hubunits{\ifmmode {\>{\rm km\ s^{-1}\ Mpc^{-1}}}\else {km
s$^{-1}$ Mpc$^{-1}$}\fi}
\def\gta{\;\lower 0.5ex\hbox{$\buildrel > \over \sim\ $}}
\def\lta{\;\lower 0.5ex\hbox{$\buildrel < \over \sim\ $}}
\def\phiIV{\ifmmode{\varphi_4}\else {$\varphi_4$}\fi}
\def\phiI{\ifmmode{\varphi_i}\else {$\varphi_i$}\fi}
\def\be{\begin{equation}}
\def\ee{\end{equation}}
\def\bea{\begin{eqnarray}}
\def\eea{\end{eqnarray}}
\def\beas{\begin{eqnarray*}}
\def\eeas{\end{eqnarray*}}
\def\gtrapprox{\;\lower 0.5ex\hbox{$\buildrel >\over \sim\ $}}
\def\lessapprox{\;\lower 0.5ex\hbox{$\buildrel < \over \sim\ $}}

\def\Em    {${\cal E}_m$}

\def\tauLL {\ifmmode{\tau_{\scriptscriptstyle LL}}\else 
           {$\tau_{\scriptscriptstyle LL}$}\fi}

\def\Em{\ifmmode{{\rm E}_m}\else {{\rm E}$_m$}\fi}
\def\NH{\ifmmode{{\rm N}_{\scriptscriptstyle\rm H}}\else {{\rm N}$_{\scriptscriptstyle\rm H}$}\fi}

\def\eg    {{\it e.g.}}

\def\etal  {\ et al.}
\def\kms{\ifmmode {\>{\rm\ km\ s}^{-1}}\else {\ km s$^{-1}$}\fi}
\def\Em{\ifmmode{{\cal E}_m}\else {{\cal E}$_m$}\fi}
\def\Dm{\ifmmode{{\cal D}_m}\else {{\cal D}$_m$}\fi}
\def\fesc{\ifmmode{\hat{f}_{\rm esc}}\else {$\hat{f}_{\rm esc}$}\fi}
\def\fescs{\ifmmode{f_{\rm esc}}\else {$f_{\rm esc}$}\fi}
\def\rsolar{\ifmmode{r_\odot}\else {$r_\odot$}\fi}
\def\emunit{\ifmmode{{\rm cm}^{-6}{\rm\ pc}}\else {
cm$^{-6}$ pc}\fi}
\def\intensity{\ifmmode{{\rm erg\ cm}^{-2}{\rm\ s}^{-1}
      {\rm\ Hz}^{-1}{\rm\ sr}^{-1}}
      \else {erg cm$^{-2}$ s$^{-1}$ Hz$^{-1}$ sr$^{-1}$}\fi}
\def\flux{\ifmmode{{\rm erg\ cm}^{-2}{\rm\ s}^{-1}}\else {erg
cm$^{-2}$ s$^{-1}$}\fi}
\def\fluxdensity{\ifmmode{{\rm erg\ cm^{-2}\ s^{-1}\ Hz^{-1}}}\else {erg
cm$^{-2}$ s$^{-1}$ Hz$^{-1}$}\fi}
\def\phoflux{\ifmmode{{\rm phot\ cm}^{-2}{\rm\ s}^{-1}}\else {phot
cm$^{-2}$ s$^{-1}$}\fi}
\def\phorate{\ifmmode{{\rm phot\ s}^{-1}}\else {phot s$^{-1}$}\fi}

\def\apjs{{\it Ap.J.Suppl.~}}

\def\etal{{\it et al.~\/}}

\def\eg{{\it e.g.}}
\def\ltsima{$\; \buildrel < \over \sim \;$}
\def\simlt{\lower.5ex\hbox{\ltsima}}
\def\gtsima{$\; \buildrel > \over \sim \;$}
\def\simgt{\lower.5ex\hbox{\gtsima}}
\def\fesc{{$\langle f_{\rm esc}\rangle$}\xspace}
\def\h2{H$_2$\xspace}

\begin{document}

\title{Density profiles of dark matter haloes on Galactic and Cluster scales}

\author{A. Del Popolo\inst{1,2,3} \& P. Kroupa\inst{2}
}
\titlerunning{Density profiles of dark matter haloes}
\authorrunning{A. Del Popolo}
\date{}
\offprints{A. Del Popolo, E-mail:antonino.delpopolo@unibg.it}
\institute{
$1$ Dipartimento di Fisica e Astronomia, Universit\'a di Catania, Viale Andrea Doria 6, 95125 Catania, Italy\\
$^2$ Argelander-Institut f\"ur Astronomie, Auf dem H\"ugel 71, D-53121 Bonn, Germany\\
$3$ Istanbul Technical University, Ayazaga Campus,  Faculty of Science and Letters,  34469 Maslak/ISTANBUL, Turkey\\
}
\abstract{
In the present paper, we improve the ``Extended Secondary Infall Model" (ESIM) of Williams et al. (2004) to obtain further
insights on the cusp/core problem. The model 
takes into account the effect of ordered and random angular momentum, dynamical friction and baryon adiabatic contraction in order to 
obtain a secondary infall model more close to the collapse reality. The model is applied to 
structures on galactic scales (normal and dwarf spiral galaxies) and on cluster of galaxies scales. 
The results obtained 
suggest that angular momentum and dynamical friction are able, on galactic scales, to overcome the competing effect of adiabatic contraction eliminating the cusp. 
The NFW profile can be reobtained, in our model only if the system is constituted just by dark matter and 
the magnitude of  angular momentum and dynamical friction are reduced 
with respect to the values predicted by the model itself.
The rotation curves of four LSB galaxies from 
de Blok \& Bosma (2002) are compared to
the rotation curves obtained by the model in the present paper obtaining a good fit to the observational data. 
On scales smaller than $\simeq 10^{11} h^{-1} M_{\odot}$ the slope $\alpha \simeq 0$ and
on cluster scales we observe a similar evolution of the dark matter density profile but in this case the density profile slope 
flattens to $\alpha \simeq 0.6$ for a cluster of $\simeq 10^{14} h^{-1} M_{\odot}$. The total mass profile, differently from that of dark matter,
shows a central cusp well fitted by a NFW model.
}

\maketitle

\section{Introduction}

The structure of dark matter haloes is of fundamental importance in
the study of the formation and evolution of galaxies and clusters of
galaxies. At the simplest level, dark matter halos form when, in the early Universe, the matter
within (and surrounding) an overdense region suffers gravitational retardation, decouples
from the Hubble flow, collapses, and in due course, virializes.
From the theoretical point of view, the structure of dark
matter haloes can be studied both analytically and numerically. A
great part of the analytical work done so far is based on the
secondary infall model (SIM) introduced by Gunn \& Gott (1972), Gott (1975) and Gunn (1977).
Calculations based on this model predict that the density profile of
the virialized halo should scale as $\rho \propto r^{-9/4}$. 
Self-similar solutions were found by Fillmore \& Goldreich (1984) and Bertschinger
(1985), while Hoffman \& Shaham (1985) (hereafter HS) studied density profiles around density peaks.
More recently modifications of  the self-similar collapse model to include more realistic dynamics of the growth process have been
proposed (e.g.  Avila-Reese et al. 1998; Nusser \& Sheth 1999; Henriksen \& Widrow 1999; Subramanian et al. 2000; Del Popolo et al. 2000 (hereafter DP2000)). 
Ryden \& Gunn (1987) (hereafter RG87) were the first  to relax the assumption of purely radial self-similar collapse by including
non-radial motions arising from secondary perturbations in the halo.
Numerous authors have emphasized the effect of an isotropic velocity dispersion (thus of non-radial
motion) in the core of collisionless haloes. 
One common result of the previous studies (RG87; White \& Zaritski 1992; Avila-Reese et al. 1998; Hiotelis 2002; Nusser 2001; Le Delliou \& Enriksen 2003; Ascasibar et al. 2004; Williams et al. 2004) is that larger amount of angular momentum leads to shallower final density profiles in the inner region of the halo. Moreover, baryons have been invoked both to shallow (El-Zant et al. 2001 (hereafter EZ01), 2003; Romano-Diaz et al. 2008)
and to steepen (Blumenthal et al. 1986) the dark matter profile.

As previously reported, the structure of dark matter haloes can also be studied through numerical simulations. Quinn, Salmon \& Zurek (1986) pioneered the use of N-body simulations to study halo formation and confirmed HS results.
More recent studies, 
(Dubinski \& Carlberg 1991, Lemson 1995, Cole \& Lacey 1996, 
Navarro et al. 1996, 1997 (NFW), Moore et al. 1998, Jing \& Suto 2000, Klypin et al. 2001, 
Bullock et al. 2001, Power et al. 2003 and Navarro et al. 2004) 
found that although the spherically-averaged density profiles of the
N-body dark matter halos are similar, regardless of the mass of the
halo or the cosmological model, their profiles are significantly different from
the single power laws predicted by the theoretical studies.
The N-body profiles are characterized by an $r^{-3}$ decline at large
radii and a cuspy profile of the form $\rho(r)\propto r^{-\alpha}$,
where $\alpha < 2$ near the center. The actual value of the inner density 
slope $\alpha$ is a matter of some controversy, with NFW 
suggesting $\alpha=1$, but with Moore et al.
(1998), Ghigna et al. (2000) and Fukushige \& Makino (2001) arguing for
$\alpha=1.5$, while Jing \& Suto (2000) and Klypin et al. (2001) claimed that the actual value of $\alpha$
may depend on halo mass, merger history, and substructure.
Power et al. (2003) pointed out that the
logarithmic slope becomes increasingly shallow inwards, with little
sign of approaching an asymptotic value at the resolved radii. In that
case, the precise value of $\alpha$, at a given cut-off scale,
would not be particularly meaningful.
This result has been later confirmed by Hayashi et al. (2003) and Fukushige et al. (2004), 
and it is predicted by several analytical
models (e.g., Taylor \& Navarro (2001), Hoeft et al. 2003).
Finally, Navarro et al. (2004) proposed a new fitting formula having a logarithmic 
slope that decreases inward more gradually than the NFW profile.

While numerical simulations universally produce a cuspy density profile,
observed rotation curves of dwarf spiral and LSB galaxies
seem to indicate that the shape of the density profile at small scales
is significantly shallower than what is found in numerical simulations
(Flores \& Primak 1994; Moore 1994; Burkert 1995; Kravtsov et al. 1998; Salucci \& Burkert 2000; Borriello \& Salucci 2001; de Blok et al. 2001; de Blok \& Bosma 2002; Marchesini et al. 2002; de Blok 2003; de Blok, Bosma \& McGaugh 2003). 
It seems that the data generally favor logarithmic density slopes close to
$0.2$ (de Blok 2003; de Blok, Bosma \& McGaugh 2003; Spekkens et al. 2005).

On cluster scales, X-ray analyses have led to wide ranging results, from $\alpha=0.6$ (Ettori et al. 2002)
to $\alpha=1.2$ (Lewis et al. 2003) or even $\alpha=1.9$ (Arabadjis et al. 2002).
Ricotti's (2003) N-body simulations suggest that density profile of DM haloes 
is not universal (in agreement with Jing \& Suto 2000; Subramanian et al. 2000; Simon et al. 2003b; Cen et al. 2004; Ricotti \& Wilkinson 2004; Ricotti et al. 2007), presenting shallower cores 
in dwarf galaxies and steeper cores in clusters. 

The discrepancy between simulations and observations has been often signaled as a genuine crisis of the CDM scenario and has become known as the ``cusp/core" problem. 
Since LSB galaxies are thought to be ideal for the comparison with theory, as their dynamics are dominated 
by dark matter with little contribution from baryons (Bothun \etal 1997), 
the discrepancy with simulations is particularly troublesome.
The significance of this disagreement, though,
remains controversial and different solutions have been proposed.
A number of authors attribute the problem to
a real failure of the CDM model, or to that of simulations (de Blok et
al. 2001a; de Blok, McGaugh, \& Rubin 2001b; Borriello \& Salucci
2001; de Blok, Bosma, \& McGaugh 2003). This has led to suggestions that dark matter properties may deviate from standard
CDM and  several alternatives have been
suggested, such as warm (Colin et al. 2000; Sommer-Larsen \& Dolgov 2001),
repulsive (Goodman 2000) fluid (Peebles 2000), fuzzy (Hu et al. 2000),
decaying (Cen 2001), annihilating (Kaplinghat et al. 2000), or self-interacting (Spergel \& Steinhardt 2000; Yoshida et al. 2000; Dave et al. 2001)
dark matter. Others argue that the inconsistency may reflect the finite resolution of
the observations that has not been properly accounted for in the analysis of
the HI rotation curves  (van den Bosch \etal\ 2000; van den Bosch \& Swaters
2001; Rhee et al. 2004). Alternatively, it has been suggested that stellar feedback from the first generation of stars formed 
in galaxies was so efficient that the remaining gas was expelled on a timescale comparable to, or less than, the local dynamical timescale. The dark matter subsequently adjusted to form an approximately constant density core (e.g., Gelato \& Sommer-Larsen 1999). This is however unlikely to affect cluster cusps.   

Simulations, observations and semi-analytic models agree on outer parts of the haloes' structure 
the disagreement in the inner regions could be connected to limits in numerical simulations (de Blok 2003; Taylor et al. 2004) or to the fact that dissipationless N-body simulations do not take into account the effects of baryons on dark matter evolution. Interestingly, the amount of central substructure seen in the semi-analytic haloes 
is consistent with the amount of substructure inferred from strong lensing experiments (Taylor et al. 2004).
Thus the semi-analytic haloes might provide a more accurate picture of the spatial distribution of substructure around galaxies even if 
an analytical method, no matter how sophisticated, will never be able to capture the full extent of complexity of a non-linear process.

Nevertheless the very large amount of work carried out by many researchers to date 
using N-body simulations has met with limited success in elucidating the
physics of halo formation. The reason is due to the fact that the  point of force of numerical simulations 
(namely to capture the full extent of complexity of a non-linear process) 
is also their weakness: numerical simulations yield little physical insight beyond empirical findings precisely because 
they are so rich in dynamical processes, which are hard to disentangle and 
interpret in terms of underlying physics.
Analytical and semi-analytical models are much more flexible than N-body simulations (see Williams et al. 2004).
So even if analytical models like SIM treat collapse and virialization of halos that are spherically symmetric, 
that have suffered no major mergers, and that have suffered quiescent accretion, 
they are worth investigating.
One of the most used semi-analytical models is the SIM whose 
most often questioned assumptions are the spherical symmetry and the absence of peculiar
velocities (non-radial motions): in the ``real" collapse, accretion does not
happen in spherical shells but by aggregation of subclumps of matter which
have already collapsed; a large fraction of observed clusters of galaxies
exhibit significant substructure (Kriessler et al. 1995). Motions are not
purely radial, especially when the perturbation detaches from the general
expansion. Nevertheless the SIM gives good results in describing the formation
of dark matter haloes, because in energy space the collapse is ordered
and gentle, differently from the chaotic collapse that is seen in N-body simulations
(Zaroubi, Naim \& Hoffman 1996). 
This is confirmed in other studies (T\'{o}th \& Ostriker 1992; Huss, Jain \& Steinmetz 1999a,b; Moore et al. 1999).
We should also add that analytical and semi-analytical models have some advantages over N-body simulations: a) they
are flexible (one can study the effects of physical processes one at a time); b) one   
can incorporate many physical effects at least in a schematic manner; c)  
they are computationally efficient (it takes about 10 s to compute the density profile of a given object at a given epoch on a desktop PC, Ascasibar et al. 2007). 

In this paper, we shall present an analytical model for haloes formation based on SIM. 
The model is an extension of the Williams et al. (2004) model to take account of ordered angular momentum, dynamical friction and adiabatic contraction. 
The model derives the initial shape of proto-halos from the fluctuation spectrum at high redshifts,
and halos are endowed with secondary perturbations, which impart random motions to halo
particles, as in Williams et al. (2004), and ordered angular momentum. The statistical properties of the secondary 
perturbations are derived from the same fluctuation power spectrum, and therefore their effects on particles random velocities
are treated self-consistently. The ordered angular momentum acquired by the proto-structures is obtained from the tidal interaction of the
proto-structure with the neighboring ones and using the theory of random fields. Dynamical friction is calculated according to Kandrup's (1980) approach. Adiabatic contraction is taken into account using the Blumenthal et al. (1986) model as modified by 
Gnedin et al. (2004).

The plan of the paper is the following: in Section 2, we introduce the model that shall be used to calculate the density profiles. 
In Section 3, we show the results of the model, while Section 4 is devoted to conclusions.

\section{Model}

As shown by the spherical collapse model\footnote{A slightly overdense sphere, embedded in the Universe, is a useful non-linear model, as it behaves exactly as a closed sub-universe because of Birkhoff's theorem. The sphere is divided into spherical ``shells". A spherical ``shell" may be defined as the set of particles at a given radius that are all at the same phase in their orbits (see Le Delliou \& Henriksen 2003).} of Gunn \& Gott (1972), a bound mass shell having initial comoving radius $x_i$ will expand to a maximum radius $x_m$ (named apapsis or turn-around radius 
$x_{ta}$). As successive shells expand to their maximum radius they acquire angular momentum and then fall in on orbits determined by the angular momentum. Dissipative physics and the process of violent relaxation will eventually intervene and convert the kinetic energy of collapse into random motions (virialization). 
Following RG87 and R88a,b  aproach, we restrict ourselves to halos that form primarily via nearly-smooth accretion of
matter, treating collapse and virialization of halos that are
spherically symmetric, that have suffered no major mergers, and that have
experienced only quiescent accretion of somewhat lumpy material.
The most important feature of the original RG87 method is that the dynamical evolution
is carried out while conserving angular and radial momenta of individual halo shells. The
initial shape of the proto-halos is derived from the fluctuation spectrum at high redshifts,
and halos are endowed with secondary perturbations, which impart random motions to halo
particles. The statistical properties of the secondary perturbations are derived from the
same fluctuation power spectrum, and therefore their effects on particles' random velocities
are treated self-consistently. Ordered angular momentum is also treated self-consistently as shown in R88a.

It is possible to divide the halo into four regions starting from the outside: in Region (1), furthest from the center
of the initial density peak, dark matter particles are beginning to feel the gravitational tug of
the central peak and are just starting to fall behind the Hubble flow. Further, in Region
(2), the particles are starting to decouple from the Hubble flow and are about to begin
collapse. In Region (3) the central density peak dominates the motion of particles; this is
the region of infall and shell crossing. Finally, in Region (4), the central part of the density
peak, virialization is taking place, or has already been reached. 
In Region (4), virialization is well underway or is complete. Many analytical solutions
in regions (1)-(3) ignore non-radial motions of particles; all orbits are assumed to be purely
radial. 
The calculations carried out in RG87 demonstrate
that non-radial motions are very important in determining the outcome in this region. Nonradial
motions give particles angular momentum, which prevents them from penetrating to
the very center of the halo (confirmed by e.g., Avila-Reese,
Firmani \& Hernandez (1998), Subramanian et al. (2000), and Hozumi, Burkert
\& Fujiwara (2000). \\

In most popular cosmological scenarios the density field soon after recombination can
be represented by a Gaussian random field. High density contrast peaks in the field will eventually achieve overdensities 
of order 1 and enter a non-linear stage of evolution. These peaks will then collapse
to form bound structures. We start with one of these peaks, and, for simplicity take it
to be spherically symmetric. The peak is divided into a very small central core and many
spherically symmetric concentric mass shells, each labeled by its initial comoving distance
from the center, $x$. 


The first step in obtaining the density profile, is to calculate the initial  
density profile arising from a primordial fluctuation, $\overline \delta_i (x_i)$. 
A well known result is the expression for the radial density profile of a fluctuation centered on a primordial 
peak of arbitrary height $\nu$:
\begin{equation}
\langle \delta (x) \rangle =\delta_0(x)=\frac{\nu \xi (x)}{\xi (0)^{1/2}}-\frac{\vartheta (\nu
\gamma ,\gamma )}{\gamma (1-\gamma ^2)}\left[ \gamma ^2\xi (x)+\frac{%
R_{\ast }^2}3\nabla ^2\xi(x) \right] \cdot \xi (0)^{-1/2} 
\label{eq:dens}
\end{equation}
(BBKS; RG87),
where $x$ is the comoving
separation, $\nu=\delta(0)/\sigma $\footnote{$\sigma$ is the mass variance filtered on a scale $R_f$} is the height of a density peak, $\xi (r)$ is the two-point 
correlation function:
\begin{equation}
\xi(x)= \frac{1}{2 \pi^2 x} \int_0^{\infty} P(k) k \sin(k x) d k
\end{equation}
$\gamma $ and $R_{\ast}$ are two spectral parameters (see BBKS) and $\vartheta (\nu \gamma ,\gamma )$ is given in BBKS.

The CDM spectrum used in this paper is that of BBKS (equation~(G3)), with transfer function:
\begin{equation}
T(k) = \frac{[\ln \left( 1+2.34 q\right)]}{2.34 q}
\cdot [1+3.89q+
(16.1 q)^2+(5.46 q)^3+(6.71)^4]^{-1/4}
\label{eq:ma5}
\end{equation}
where 
$q=\frac{k\theta^{1/2}}{\Omega_{\rm X} h^2 {\rm Mpc^{-1}}}$.
Here $\theta=\rho_{\rm er}/(1.68 \rho_{\rm \gamma})$
represents the ratio of the energy density in relativistic particles to
that in photons ($\theta=1$ corresponds to photons and three flavors of
relativistic neutrinos). The spectrum is connected to the transfer function through the equation:
\begin{equation}
P(k)=P_{CDM} e^{-1/2 k^2 R_f^2}
\end{equation}
where $R_f$ is the smoothing (filtering) scale and $P_{CDM}$ is given by:
\begin{equation}
P_{CDM}= A k T^2(k)
\end{equation}
where $A$ is the normalization constant. 
We normalized the spectrum by imposing that the mass variance of the density field
\begin{equation}
\sigma^2(M)=\frac{1}{2 \pi^2} \int_0^\infty dk k^2 P(k) W^2(kR)
\end{equation}
convolved with the top hat window 
\begin{equation}
W(kR)=\frac{3}{(kR)^3} (\sin kR-kR \cos kR)
\end{equation}
of radius 8 $h^{-1}$ $Mpc^{-1}$ is $\sigma _{8}=0.76$ (Romano-Diaz et al. 2008).
Throughout the paper we adopt a 
$\Lambda$CDM cosmology with WMAP3 parameters, $\Omega_m=1-\Omega_{\Lambda}=0.24$,  $\Omega_{\Lambda}=0.76$, $\Omega_b=0.043$ and $h=0.73$, where $h$ is the Hubble constant in units of 100 km $s^{-1}$ $Mpc^{-1}$. 
The density profile of an initial, pre-collapse halo, at early times, $\delta_i(x)$, is related to 
$\delta_0(x)$ by the linear growth factor $D(z)$ (Peebles 1980) by:
\begin{equation}
\delta_i(x)=\delta_0(x)/D(z_i)
\end{equation}
Let us divide the main smooth spherically symmetric halo, $\delta(x)$, into many concentric
mass shells. Each shell is uniquely labeled by $x$, its initial comoving radius. The halo
represents an upward departure from average background density, i.e. density interior to
any shell is greater than critical at all times. 
Each shell's evolution is divided into two stages. Initially a shell expands
with the Hubble flow, but with a slight deceleration arising from the central mass concentration.
Eventually the shell's outward radial velocity decreases to zero, after which the shell
collapses, by some distance, back towards the center of the halo. The dividing moment is
called the turn-around, and at any given time corresponds to the line dividing Regions (1)
and (2). In a halo with a declining density profile the turn-around happens at progressively
later cosmic times for shells at greater distances from the center.

In such cases the time evolution of a shell until
it reaches turn-around is given by a set of parametric equations (Gunn \& Gott 1972, Peebles
1980),
\begin{equation}
r(\theta)={1\over 2}x{\bar\delta_0}^{-1}(1-\cos\theta),
\label{rtheta}
\end{equation}
\begin{equation}
t(\theta)={3\over 4}t_0{\bar\delta_0}^{-3/2}(\theta-\sin\theta),
\label{ttheta}
\end{equation}
where $r$ is particle's proper radius, $t$ is cosmic time, and $\bar\delta_0(x)$
is the initial average fractional density excess inside the shell
(Eq. [13] of RG87), 
\begin{equation}
\bar\delta_0(x)={3\over x^2}\; \int_0^x \delta_0(y)\;y^2\;dy,
\label{eq:overd}
\end{equation}
and $t_0$ is the present time. 
The mass of a shell, and the mass within a shell are constant; 
these relations, together with Eqs.~\ref{rtheta} and \ref{ttheta} can be used
to compute the fractional overdensity for any shell, parameterized by its
$\delta_0/{\bar\delta_0}$, at a cosmic time corresponding to $\theta$, 
(Eq. [23] of RG87):
\begin{equation}
\delta(\theta)+1={{9(\theta-\sin\theta)^2}\over{2(1-\cos\theta)^3}}
\Biggl(1+3\Bigl[1-{{\delta_0}\over{\bar\delta_0}}\Bigr]
\Bigl[1-{{3\sin\theta(\theta-\sin\theta)}\over{2(1-\cos\theta)^2}}\Bigr]
\Biggr)^{-1},
\label{deltatheta}
\end{equation}
This can be used to construct the density run of a halo at a fixed cosmic 
time $t/t_0$.


The mean density distribution about a peak is spherically symmetric. However, the
initial density peak will in general have a triaxial shape, leading to non-spherical collapse.
As shown by R88a, since the quadrupole moment of the protogalaxy is largely due to the outermost shells, where triaxiality is insignificant, it is justifiable to ignore the triaxility in computations. In other terms,
the dynamics of the halo collapse are dictated by the potential, which, being a double
integral over all space, is much rounder than the mass distribution. Therefore the effects
of intrinsic triaxiality of initial density peaks are smaller than those due to the secondary
perturbations, and so can be ignored. Furthermore, initial triaxiality is less severe in larger,
2-4 $\sigma$ peaks (Bardeen et al. 1986), which are the subject of the present study. With this caveat
the smooth part of our density peak is still described by Eq. (\ref{eq:dens}). 

Moreover, in reality, the initial density peak will not be smooth, but will instead be sprinkled with
many smaller scale positive and negative perturbations that arise from the same Gaussian
random field that gave rise to the main peak. These secondary perturbations will perturb
the motion of the dark matter particles from their otherwise purely radial orbits.

So, in order to investigate effects such as
tidal torques and nonradial collapse, it is necessary to consider the nonspherical portion of the density distribution.
In addition to the smooth halo, RG87 and Williams et al. (2004) considered contributions from the secondary 
perturbations which arise from the same Gaussian random field that gave rise
to the main halo. 
The overall initial density profile, linearly evolved to the present day, 
can be written as,
\begin{equation}
\rho({\bf x})=\rho_0[1+\delta_0(x)][1+\epsilon_0({\bf x})],\label{den}
\end{equation}
where $\rho_0$ is the present day background density, density excess due to
the main halo is $\delta_0(x)$, and is assumed to be spherically symmetric, 
and $\epsilon_0({\bf x})$ is the density excess contributed by the random 
secondary perturbations. 
In a statistical sense, the growth rate will depend on $\delta_0/\bar\delta_0$,
or, equivalently, $x$, either of which can be used to parameterize the strength 
of the tidal field in a spherically symmetric halo. 
The final expression for the 
growth rate is given by (Eq. [28] of RG87 and Williams et al. 2004)
\begin{equation}
\epsilon(x,\theta)={\epsilon_0(x)\over{\bar\delta_0}}
{{f_2(\theta)}\over{f_1(\theta)-[\delta_0(x)/\bar\delta_0(x)]f_2(\theta)}},
\label{epsilontheta}
\end{equation}
where 
$\epsilon_0(x)$ is the amplitude of the initial perturbation, 
given by 
\begin{equation}
\epsilon_0({\bf x})={1\over{(2\pi)^3}} 
\int d^3k\; \epsilon_{\bf k}\;e^{i \bf k\cdot \bf x},\quad \quad
\langle |\epsilon_{\bf k}|^2 \rangle =P(k)
\label{epsilonfield}
\end{equation}
and 
$f_1$ and $f_2$ are functions of the `time' parameter $\theta$:
$f_1(\theta)=16-16\cos\theta+\sin^2\theta-9\theta\sin\theta$, and
$f_2(\theta)=12-12\cos\theta+3\sin^2\theta-9\theta\sin\theta$.

At any given time and proper position the random acceleration due to 
perturbation field $\epsilon({\bf r})$ is given by (RG87 Eq. [40]): 
\begin{equation}
{\bf g}({\bf r},t)={\bf g}_{tot}({\bf r},t)-{\bf g}_b({\bf r},t)\approx
G\int d^3r^\prime
{{\rho_b ({\bf r^\prime},t)\epsilon({\bf r},t)}\over
{|{\bf r^\prime}-{\bf r}|^3}}({\bf r^\prime}-{\bf r}),
\label{accln}
\end{equation}
which is an integral over all space, and 
$\rho_b({\bf r},t)=\rho_0 (t) [1+\delta({\bf r},t)]$ is the background density due
to the main halo, and is related to the total density given by Eq.~\ref{den} by, 
$\rho({\bf r},t)=\rho_b({\bf r},t)[1+\epsilon({\bf r},t)]$; $\rho_0 (t)$ is the
average density of the Universe at epoch $t$.
Remembering that the density distribution in the main peak and perturbation 
field and growth rate of perturbations are functions of radial position only, we 
use $x$ instead of ${\bf x}$, and $r$ instead of ${\bf r}$.
Since, our goal is the {\it rms} value of acceleration, so scalar $g$ 
will replace vector {\bf g}. 
A major simplification is accomplished by decoupling the time dependence of 
acceleration, i.e. the rate of growth of acceleration, from its spatial variation.
Let the initial acceleration field due to secondary perturbations be denoted by
$g_0(x)$. Then the dimensionless rate of growth of acceleration is given by,
\begin{equation}
F_g(x,t)=g(x,t)/g_0(x)=g(r,t)/g_0(r).
\label{Fg_def}
\end{equation}
With these, the proper displacement of a particle can be evaluated as 
\begin{equation}
d_p(x,t)=\int_0^t dt_1 \int_0^{t_1} dt_2\; g(x,t_2)
=g_0(x)\int_0^t dt_1 \int_0^{t_1} dt_2\; F_g(x,t_2)= \Delta g_0(x) t_0^2 F_r(x,t),
\label{Fg}
\end{equation}
Next we describe the two functions, $F_g(x,t)$ and $g_0(x)$, separately.

The
acceleration growth rate becomes 
(RG87 Eq. [44]\footnote{Note a typo in their paper: 
$f_1(\theta)$ in the numerator of Eq. (44) should be $f_2(\theta)$.})
\begin{equation}
F_g(x,\theta)={g(x,\theta)\over g_0(x)}= 
{{\rho_0[1+\delta(x,\theta)]\epsilon(x,\theta)r(x,\theta)}
\over{\rho_0\epsilon_0 x}}=
8\bar\delta_0{{f_2(\theta)}\over
{[f_1(\theta)-{{\delta_0(x)}\over{\bar\delta_0(x)}}\;f_2(\theta)]^2}},
\label{approx_accln}
\end{equation}

The spatial dependence of acceleration is given by:
\begin{equation}
\Delta g_0(d,x)=
4 G \rho_0
\Bigl[
\int P_s(k)e^{-kd}(1-{{\sin kx}\over{kx}})dk\Bigr]^{1/2}.
\label{deltag0}
\end{equation}

The comoving displacement is given by:
\begin{equation}
d(x,t)=d_p(x,t)/a(t)=\Delta g_0(x) t_0^2 F_r(x,t)/a(t).
\label{Fgg}
\end{equation}

In the early part of the evolution of the halo most of the shells are still expanding. During
this time secondary perturbations grow, and so does the acceleration, the velocity, and the
displacement contributed by these perturbations to the particles in the shell. Because the
secondary peaks are randomly distributed within the halo, they displace the dark matter
particles in random directions from their original positions. This can be visualized as a shell
having an internal velocity dispersion, resulting in a ``swollen" shell. 
If we concentrate on a single particle, its orbit, viewed from the rest frame of the parent shell, will oscillate between
an inner and an outer radius of that shell (i.e. pericenter, $r_p$ and apocenter, $r_a$). In general, this orbit
will not be closed and will resemble a rosette. 
In a real situation, the ``swell" of any given shell will gradually increase as the shell
expands away from the center of the halo, and the influence of the secondary peaks grows, 
but for simplicity, as long as a given shell is expanding its dark matter particle positions and velocities are not
corrected for the effects of secondary perturbations.
So, the contributions are evaluated analytically but are not imparted to the shell until it reaches turnaround.

The magnitude of the extra velocity imparted to a typical dark matter particle
at the time of turn-around (Eq. [48] of RG87),
\begin{equation}
|\Delta {\bf v_{\rm rms}}(x,t_c/2)|=F_v(x,t_c/2)\Delta g_0[d(x,t_c/2),x]t_0,
\label{deltav}
\end{equation}
where the velocity growth factor $F_v(x,t)$ is: 
\begin{equation}
F_v(x,t)= \int_0^\theta  F_g(x,\theta_1) \frac{dt}{d \theta_1} d \theta_1
\end{equation}
and the tangential, $v_{tan}$, and radial, $v_{rad}$, components of velocity are:
\begin{equation}
(\Delta v_{tan})^2={2\over 3}|\Delta {\bf v_{\rm rms}}(x)|^2,\quad {\rm and} \quad
(\Delta v_{rad})^2={1\over 3}|\Delta {\bf v_{\rm rms}}(x)|^2. 
\label{deltav_vec}
\end{equation}
The mean radial velocity at $t=t_c/2$ is zero, while the mean tangential velocity is given by:
\begin{equation}
j_\theta(x)=\left (\frac{2}{3} \right)^{1/2} r_m \Delta v(x,t_c/2)
\end{equation}
It is important to note that $j_\theta$ is inversely correlated with the height of the peak in $\delta$, since 
both $\Delta v$ and $r_m$ (the maximum expansion radius) decrease with an increase in the height $\nu$ of the peak.
The random angular momentum is important for the particle (shell) orbits since the larger
it is, the larger is the orbital ellipticity, and then the shell penetrates less to the center,
resulting in a flattening of the inner density profile. In several of the previously quoted 
papers (even Williams et al. 2004) only the random angular momentum
was taken into consideration, while we take also account of ordered angular momentum, dynamical friction and adiabatic contraction.

The collapse starts from the inner most shell, the one adjacent to the core. 
When the first shell reaches its $r_{m}$ it collapses and finds its 
$r_{a}$ and $r_{p}$ within the overall halo potential. It is assumed that
the potential is changing slowly compared to the dynamical timescales of the
shells, so that every shell conserves its adiabatic invariants, the radial and 
tangential momenta,
\begin{eqnarray}
j_\theta(x)=\Delta v_{tan}\; r_{m}\\
\label{jtheta}
j_r(x)=\int_{r_{p}}^{r_{a}} v_{rad}\, dr
\label{jr}
\end{eqnarray}
throughout the collapse. This is an important assumption in the RG87 formalism,
it is crucial to the computation of dynamics of shell crossing.
Up to the moment when a given shell reaches its maximum expansion radius
$r_{m}(x)$ at a time $t=t_c(x)/2$ corresponding to $\theta=\pi$, the shell
is assumed to be thin, its radial extent determined by the initial shell 
separation. At $r_{m}$, the average dark matter particle in the shell is 
given its additional random velocity, Eq.~\ref{deltav} and 
Eq.~\ref{deltav_vec}. 
In either case the apocenter and pericenter can be calculated from the 
particle's energy integral and $\Delta v_{tan}$ and $\Delta v_{rad}$.
The energy integral,
\begin{equation}
E= \psi(r_m)-\frac{1}{2} \left[\left(\frac{dr}{dt}\right)^2+\left(\frac{j_\theta}{r}\right)^2 \right]
\end{equation}
%
is conserved if $\psi$ is not too rapidly changing with time, an approximation which is good outside the immediate region of the core.


The radial velocity of a particle is
\begin{equation}
v_{rad}=[2(E-\psi(r))-(j_\theta/r)^2]^{1/2},
\label{vrad}
\end{equation}
(RG87).
Note that RG87 define their
energy integral and potential as the negatives of the conventional definitions
of these quantities. We use the conventional definitions, i.e. potential is a
negative quantity inside the halo, and energy integral of a bound particle
is negative.

The addition of a random radial 
velocity ensures that the apocenter distance is greater than $r_{m}$, the radius of maximum expansion in the absence of velocity 
perturbations. A measure of the mean radial momentum of the particle is the radial action, (RG87):
\begin{equation}
j_r= \int_{r_{p}}^{r_{a}} v_r dr= \overline{v}_r dr(r_{a}-r_{p})
\end{equation}


The radial distribution
of mass within a shell between apocenter and pericenter is not uniform. The
density in the radial range $dr$ around $r$ is proportional to the amount of 
time the particle spends there, (Eq. [53] of RG87),
\begin{equation}
P(r)\;dr={{v_{rad}^{-1}\;dr}\over{\int_{r_{p}}^{r_{a}}v_{rad}^{-1}\;dr}}.
\label{probability}
\end{equation}

If $M_{shell}$ is the mass of the added shell, then the total mass distribution of the core and shell together is:
\begin{equation}
M_1(r)=M(r)+M_{shell} \int_{r_p}^r P(r') d r'
\end{equation}

From this mass function, the new potential $\psi_1(r)$ is calculated. The next mass shell is added and its probability distribution 
is calculated in the potential $\psi_1(r)$. However, the mass distribution of a newly added shell overlaps with shells which have fallen earlier; that is, the pericenter of the shell is at a smaller radius than the apocenters of some fraction of the previously added shells. Thus we must recompute after adding each new shell the orbits for each shell with which it overlaps. Since the potential is not violently changing, the adiabatic invariants of the orbits are conserved. In this case, the adiabatic invariants are the angular momentum $j_{\theta}$ and the radial action $j_r$. By
repeatedly adding shells in this manner, while adjusting the orbits so that the angular momentum and radial action of the orbits are conserved, a self-consistent mass distribution is built up.
\footnote{In other words, the mass particles, once position and velocities are assigned, are allowed to follow the appropriate orbit in the gravitational
potential of the previously collapsed matter. As each mass shell is added, those previously added shells with which it overlaps have their orbits adjusted so that the angular momentum $j_{\theta}$ and radial moments $j_r$ integrated from pericenter to apocenter, are conserved.}

%


The collapse of a perturbation 
taking into account angular momentum, and dynamical friction 
may be calculated by solving the equation for the radial acceleration (Kashlinsky 1986, 1987;
Antonuccio-Delogu \& Colafrancesco 1994; Peebles 1993):

\begin{equation}
\frac{dv_r}{dt}=\frac{h^2(r,\nu )+j^2(r, \nu)}{r^3}-G(r) -\mu \frac{dr}{dt}
\label{eq:coll}
\end{equation}

where $h(r,\nu )$ 
is the ordered specific angular momentum generated by tidal torques, $j(r, \nu)$ the random angular momentum (see RG87), 
$G(r)$ the acceleration, 
and $\mu$ the coefficient of dynamical friction.

In the peculiar case of $\mu=0$, Eq. (\ref{eq:coll}) can be integrated to obtain the square of velocity:
\begin{equation}
v(r)^2=2 \left[\epsilon -G \int_0^r \frac{m_T(y)}{y^2} d y +\int_0^r \frac{h^2}{y^3} dy 
\right]
\end{equation}
where $\epsilon$ is the specific binding energy of the shell that can be obtained from the previous equation at turn-around when, $dr/dt=0$. 

If $\mu\neq 0$, the previous equation must be substituted with: 
\begin{equation}
\frac{d v^2}{d t}+2 \mu v^2=2 \left[
\frac{h^2+j^2}{r^3} -G\frac{m_T}{r^2} 
\right] v
\label{eq:veloc}
\end{equation}
which can be solved numerically for $v$.
In the previous equation, we have two unknown quantities, the specific ordered angular momentum, $h$, and the 
coefficient of dynamical friction, $\mu$. 

The ordered angular momentum is connected to the tidal interaction of the proto-structure with the neighboring ones.
The explanation of galaxies spins gain through tidal torques was pioneered by Hoyle
(1949). Peebles (1969) performed the first detailed calculation of the acquisition of angular
momentum in the early stages of protogalactic evolution. More recent analytic computations
(White 1984, Hoffman 1986, R88; Eisenstein \& Loeb 1995; Catelan \& Theuns 1996)
and numerical simulations (Barnes \& Efstathiou 1987) have re-investigated the role of tidal
torques in originating the angular momenta of galaxies.

In the present paper, we take into account both types of angular momentum: random $j$, and ordered, $h$. 
As described in RG87, to calculate the ordered angular momentum, one has first to 
obtain the rms torque, $\tau (r)$, on a mass shell  and then 
calculate the total specific angular momentum, $h(r,\nu )$, acquired during
expansion by integrating the torque over time (Ryden 1988a (hereafter R88), Eq. 35): 
\begin{equation}
h(r,\nu )=\frac 13\left( \frac 34\right) ^{2/3} 
\frac{\tau _ot_0}{M_{sh}}
\overline{\delta }_o^{-5/2}\int_0^\pi \frac{\left( 1-\cos \theta \right) ^3}{%
\left( \vartheta -\sin \vartheta \right) ^{4/3}}\frac{f_2(\vartheta )}{%
f_1(\vartheta )-f_2(\vartheta )\frac{\delta _o}{\overline{\delta _o}}}%
d\vartheta   \label{eq:ang}
\end{equation}
where $M_{sh}= 4 \pi \rho_{\rm b}\left[1+\delta(x)\right] x^2 \delta x$ is the mass in a thin spherical shell of internal radius $x$, $\tau_o$ is the tidal torque at time $t_0$, 
while the mean over-density inside the shell, $\overline{%
\delta }(r)$, is given by Eq. (\ref{eq:overd}).

As remarked by Del Popolo et al. (2001) the angular momentum obtained from Eq. (\ref{eq:ang}) is evaluated at the time of maximum expansion $t_{\rm m}$.
With the BBKS power spectrum (filtered on a galactic scale), 
for a $\nu=3$ peak of mass $\simeq 2 \times 10^{11} M_{\odot}$, the model gives a value of
$h=2.5 \times 10^{74}$ $\frac{g cm^2}{s}$.

It is interesting to compare this result with a different method like that of Catelan \& Theuns (1996), who calculated the
angular momentum at maximum expansion time (see their Eqs. (31)-(32)) and compared it with previous theoretical
and observational estimates. Assuming the same value of mass and $\nu$ used to obtain our previously quoted result 
and the same distribution of angular momentum 
as adopted by Catelan \& Theuns (1996), we obtain a value for the
angular momentum of $h=2.0 \times 10^{74}$ $\frac{g cm^2}{s}$ in agreement with our result.

The total angular momentum of a system is often expressed in terms of the dimensionless spin parameter,
\begin{equation}
\lambda=\frac{L |E|^{1/2}}{GM^{5/2}}, 
\end{equation}
where $L$ is the angular momentum, summed over shells, and $E$ is the binding energy of the halo. 
Expressing the values of angular momentum calculated as previously reported in terms of the spin parameter, we get a
value of $\lambda \simeq 0.04$ for a 3$\sigma$ peak of $\simeq 10^{12} M_{\odot}$\footnote{Notice that there is a very mild dependence of $\lambda$ on peak height, or mass}. This is comparable with values found in the literature
(Barnes \& Efstathiou 1987; Vivtska et al. 2002). Using the parameters in Vivitska et al. (2002), the maximum of the distribution of $\lambda$ (well approximated by a log-normal distribution) is $\lambda=0.035$ while there is a 90\% probability that $\lambda$ is in the range 0.02-0.1. 

Since our paper deals also with LSB galaxies, we must recall that LSB galaxies are more angular momentum dominated compared to normal galaxies with the same luminosity (McGaugh \& de Blok 1998). These galaxies are characterized by high values of 
$\lambda$: as shown in Boissier et al. (2002), 35\% of all galaxies (in number) having $0.06<\lambda<0.21$ are LSBs.
Therefore one expects according to several analytical papers (Nusser 2001; Hiotelis 2002; Le Delliou \& Henriksen 2003; Ascasibar et al. 2004; Williams et al. 2004) that these objects should typically have shallower density cusps. Moreover the cores of these galaxies are much less dense than what the simulations indicate. 
In the case of LSB galaxies, we assumed a value of $\lambda=0.06$ performing a conservative estimate.  
In agreement with what is reported, larger values of $\lambda$ should imply a further flattening of the rotation curves in Fig. 3. 



A more complete description of this calculation can be found in Del Popolo (2006).

It is important to note that after turn-around, not only the random angular momentum,
$j$, will contribute to originate non-radial motions in the protostructure but also the ordered
angular momentum, $h$. In fact, as successive shells expand to their maximum radius they
acquire angular momentum and then fall in on orbits determined by the angular momentum.
Actual peaks are not spherical; thus the infall of matter will not be purely radial. Random
substructure in the region surrounding the peak will divert infalling matter onto non-radial
orbits. The role of these random motions is of fundamental importance in structure
formation.

We took into account dynamical friction by introducing the dynamical friction force in the equation of motions. In ordered to calculate $\mu$, we recall that in hierarchical universes, matter is concentrated in lumps, and the lumps into groups and so on, which act as gravitational field generators. One can calculate the stochastic force generated by these field generators and then following Kandrup (1980) the dynamical friction force per unit mass
\begin{equation}
\mathcal{F}=-\mu v=-\frac{4.44[Gm_an_{ac}]^{1/2}}N\log \left( 1.12N^{2/3}\right)
\frac v{a^{3/2}}=-\beta_o \frac v{a^{3/2}}
\label{eq:dynfric}
\end{equation}
where $N=\frac{4\pi }3R_{sys}^3n_a$, is the total number of field particles, $R_{sys}$ the system radius,
$m_a$ and $n_a$ are respectively the average mass and the number density of the field particles,
$n_{ac}=n_a\times a^3$ is the comoving number of field particles and $a$ the expansion parameter, 
connected to the proper radius of a shell through: 
\begin{equation}
r(r_i,t)=r_ia(r_i,t)
\end{equation}

The number and mass of the field generators is then calculated using the theory of Gaussian random fields (BBKS). 
The acceleration obtained from Eq. (\ref{eq:dynfric}) is used in the equation of motion. 
Again, the reader is referred to Del Popolo (2006) for a more complete description.

The shape of the central density profile is influenced by baryonic collapse: baryons drag dark matter in, causing the so called adiabatic contraction (AC) steepening the dark matter density slope.
By considering a spherically symmetric protostructure, before dissipation, consisting of a baryonic fraction $F$ 
with $F << 1$ of dissipational baryons and a fraction $1-F$ of dissipationless dark matter particles constituting the halo, Blumenthal et al. (1986) 
described an approximate analytical  model to calculate the effects of AC. 
One assumes that the dissipational baryons and the halo particles are well mixed initially (i.e., the ratio of their densities is $F$ through 
the protostructure). 
This is due to the fact that the original angular momentum of the dark matter halo comes from gravitational (tidal) interactions 
with its environment. Thus, the dark matter and the gas experience the same torque in the process of halo assembly and should initially have (almost) the same specific angular momentum (Klypin et al. 2002).  
So, a usual assumption is that initially baryons had the same density profile as the dark matter (see the previous discussion and Mo et al. 1998; Cardone \& Sereno 2005; Treu \& Koopmans 2002; Keeton 2001). 

 %
%
In real systems, baryons assume a distribution in the central regions of the halo which is determined by the competition between dissipation and star formation. The dissipative infall of baryons will cease either when most of the baryon gas is converted to stars, ending dissipation, or when a rotationally supported gas disk is formed. As the baryons dissipate energy, they fall toward the center of the halo. If the baryons conserve their angular momentum as they fall inward, then the baryons will dissipate energy until they form a rotationally supported disk, whose shape is determined by the angular momentum distribution $h(M)$ of the baryons. 
So, summarizing, the two methods of ending dissipative infall result in a spheroidal distribution (if star formation ends the infall) or in a disk of stars (if angular momentum ends the infall) (R88a). 
So following the usual practice, the final baryon distribution is assumed to be a disk (for spiral galaxies) (Blumenthal
et al. 1986; Flores et al. 1993; Mo et al. 1998; Klypin et al. 2002; Cardone \& Sereno 2005).
In our calculations, we shall assume the Klypin et al (2002) model for the baryon distribution (see
their subsection 2.1), when dealing with mass scales typical of spiral galaxies. In the case
of elliptical galaxies and clusters a typical assumption (that we shall use in the paper) is
that baryons collapse to a Hernquist configuration (Rix et al. 1997; Keeton 2001; Treu \&
Koopmans 2002).
One objection that one could advance is that clearly if the final baryon distribution is assumed to be a disk, the inner gravity potential is clearly non-spherical. The usual assumption that is made at this step (see RG87; R88a,b; Mo et al. 1998; Klypin et al. 2001; Cardone et al. 2006)  is that the final configuration (in this case the disk) is assembled slowly, so that we can assume that the halo responds adiabatically to the modification of the gravitational potential and it remains spherical while contracting. The angular momentum of the dark matter particles is then conserved and a particle which is initially at radius $r$ ends up at a rdius $r'$, as described in the following.
%


The scale of the Hernquist profile is fixed by the competition between dissipation and star formation, when most of the baryon gas is converted to stars (see RG87; Gnedin et al. 2004).

We take into account adiabatic compression as done in RG87 improving it as done by Gnedin et al. (2004) (see the following). 
The Blumenthal et al. (1986), RG87, R88ab model for the adiabatic contraction can be described as follows.  
Since the baryon fraction $F$ is much less than one, the adiabatic invariants of the dark matter orbits are conserved (Blumenthal et al. 1986; RG87). If the matter is on circular orbits, the invariant is $r M(r)$; if the dark matter is on radial orbits in a self-similar distribution, the invariant is $r_m M(r_m)$ (Blumenthal et al. 1986). Since the actual orbits are neither circular non radial, the two adiabatic invariants which are preserved are the angular momentum $j_\theta$ and $j_r$, of the orbits (RG87). In the absence of dissipation, the mass distribution $M_0(r)$ has a value of $F$ which is constant throughout the structure. A self-consistent final solution, in which the baryons form a disk (or Hernquist configuration), conserving $h$, and the dark matter is compressed, conserving $j_\theta$ and $j_r$, is found by an iterative procedure. In the zeroth-order approximation, the baryons have their predissipation distribution $M_{B0}=F M_0(r)$, and the dark matter has the distribution $M_{D0}=(1-F) M_0(r)$. While the dark matter distribution is held constant, the baryons fall inward, preserving their angular momentum until they are on circular orbits. Baryons originally at radius $r$ and with specific angular momentum $h$ will end up, in the stiff halo, at a radius,
\begin{equation}
r'=\frac{h^2}{G[M_{B0}(r)+M_{D0}(r')]},
\end{equation}
The baryons now have a new, more compact distribution $M_{B1}(r)$, and the central concentration of the baryons will draw the dark matter inward. The new potential of the baryons plus the dark matter is given by,
\begin{equation}
\Phi_1(r)=-G \int_r^\infty \frac{M_{D0}(s)+M_{B1}(s)}{s^2} ds,
\end{equation}
After fixing the value of $j_\theta$ for each mass shell of the halo, the value of the apocenter is adjusted until the orbits of the dark matter in the potential $\Phi_1(r)$ have the same value of $j_r$, which they had in the predissipation distribution.
The distribution of mass $M_{D1}(r)$ of the dark matter in the potential $\Phi_1$ is built orbit by orbit, ensuring that $j_\theta$ and $j_r$ are conserved. In the second iteration one has:
\begin{equation}
r''=\frac{h^2}{G[M_{B1}(r)+M_{D1}(r'')]},
\end{equation}
and the new potential $\Phi_2$ is calculated and the dark matter orbits are adjusted to preserve $\j_\theta$ and $j_r$. The iteration 
continues till convergence. The previous model can be improved in two ways, as shown by Gnedin et al. (2004) and Klypin et al. (2002).
It is worth stressing that there is some debate of the validity about the adiabatic compression formalism, with authors like Jesseit et al. (2002) finding a substantial agreement between the final dark matter distribution in numerically simulated haloes and that predicted by the adiabatic compression approach. On the other hand, this result has been contradicted on the basis of a set of higher resolution numerical simulations by Gnedin et al. (2004). According to these authors, the standard adiabatic compression formalism systematically overpredicts the dark matter density profile in the inner 5\% of the virial radius, and the adiabatic compression formalism overpredicts the dark matter density less than $\simeq 10\%$ at $r/r_v$ $\simeq 0.1$, while the error quickly decreases for larger values of $r/r_v$.
Gustafsson et al. (2006) confirmed that the Blumenthal et al. (1986) model overestimates the central dark matter density.
Moreover they showed that the modified model proposed by Gnedin et al. (2004) even if it is a considerable
improvement it is not perfect. Moreover it is found that the contraction parameters in their model
not only depend on the orbital structure of the dark-matter–-only halos, but also on the stellar feedback
prescription which is most relevant for the baryonic distribution.
As a caveat we want to recall that according to Romano-Diaz et al. (2008), Gustafsson et al. (2006), even if focused on the AC, had an insufficient resolution to reach conclusive results. In the following, we shall take into account the Gnedin et al. (2004) simple modification which describes numerical results more accurately than Blumenthal et al. (2006).
They proposed a modified adiabatic
contraction model based on conservation of the product of the current
radius and the mass enclosed within the orbit-averaged radius,
\begin{equation}
  M(\bar{r})r= {\rm const},
  \label{eq:modified}
\end{equation}
where the orbit-averaged radius is
\begin{equation}
  \bar{r} = {2 \over T_r} \int_{r_p}^{r_a} r \, {dr \over v_r},
\end{equation}
where $T_r$ is the radial period, $r_a$ is the apocenter radius and $r_p$ the pericenter radius.  
The previous, classical AC model, assumes no angular momentum exchange between different components (e.g., 
baryons and dark matter), transfer of angular momentum from the baryons to the dark matter can be produced by 
dynamical friction, and the change of the AC model due to this effect can be taken into account by means of 
a simple model by Klypin et al. (2002).

\begin{figure}[ht]
\centerline{\hbox{(a)
\psfig{file=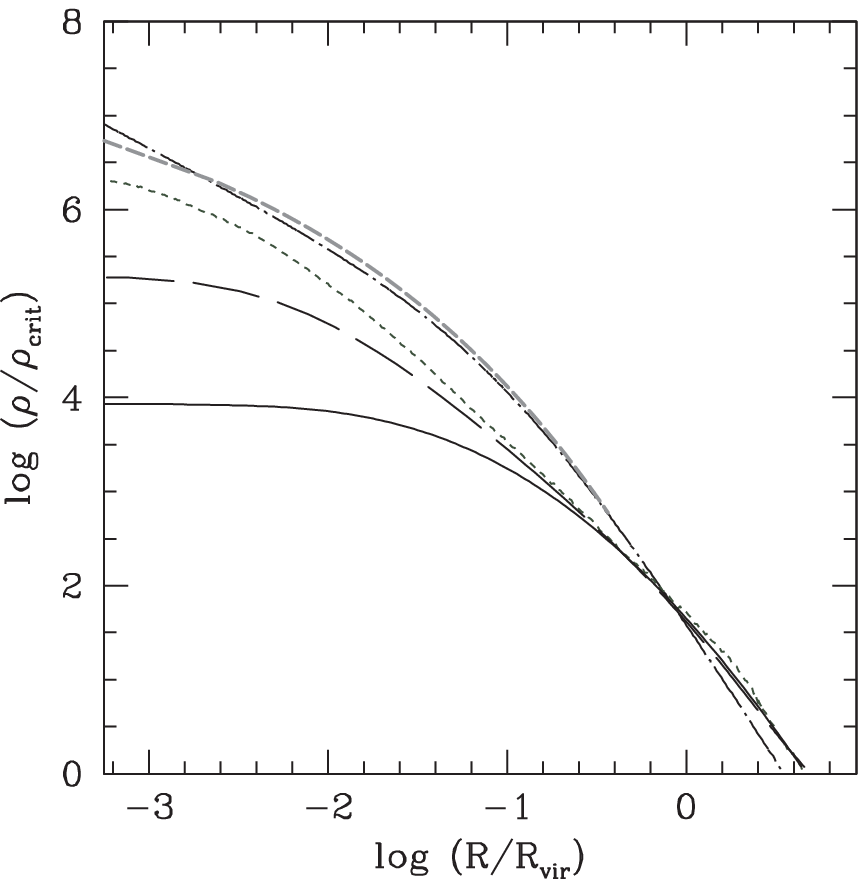,width=9.0cm}
}}
\centerline{\hbox{(b)
\psfig{file=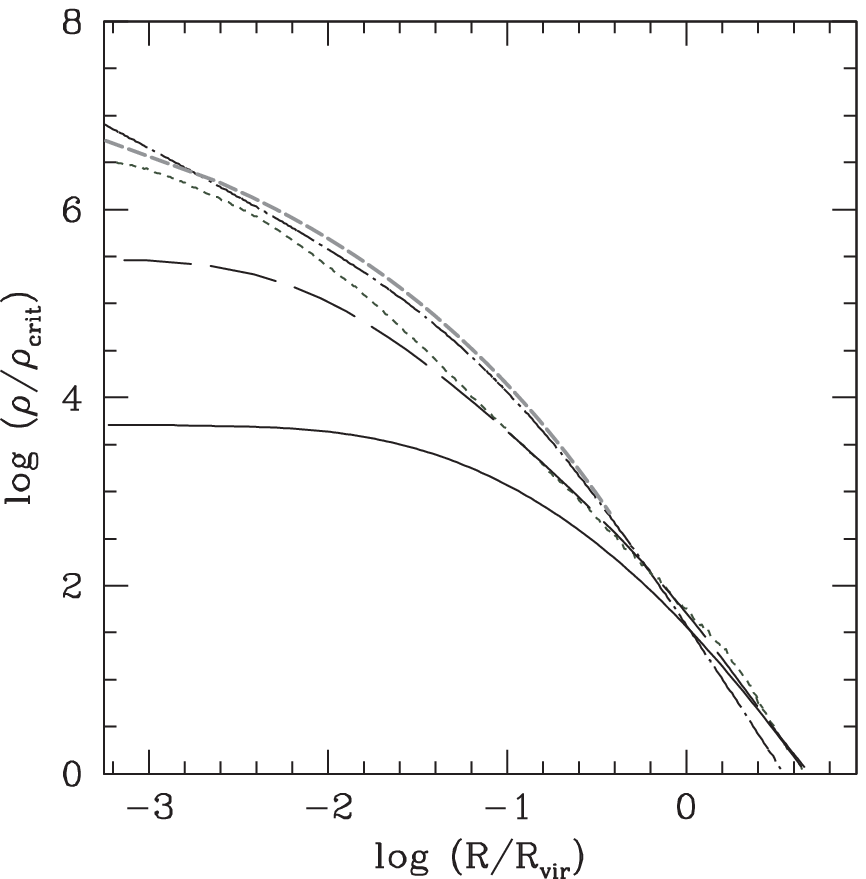,width=9.0cm}
}}
\caption{Panel (a). Dark matter haloes generated with the model of Section 2. In this case we do not take into account baryon collapse. The upper dashed-line represents the results of the Aquarius Experiment for a halo of $M \simeq 10^{12} h^{-1} M_{\odot}$, the dot-dashed line represents the NFW profile for $c=10$, 
calculated by means of Eqs. (\ref{eq:cvirr}),(\ref{eq:navar}), (\ref{eq:navarr}) expressing the scaling radius $r_s$ in terms of the virial radius.
The short-dashed line, dashed line and solid line represent, respectively, the density profile obtained by means of the model of the present paper for
$M \simeq 10^{12} h^{-1} M_{\odot}$, $M \simeq 10^{10} h^{-1} M_{\odot}$, and $M \simeq 10^{8} h^{-1} M_{\odot}$  without inclusion of baryonic infall. Panel (b) same of panel (a) but taking account of baryonic infall.
}
%
\end{figure}

\begin{figure}[ht]
\psfig{file=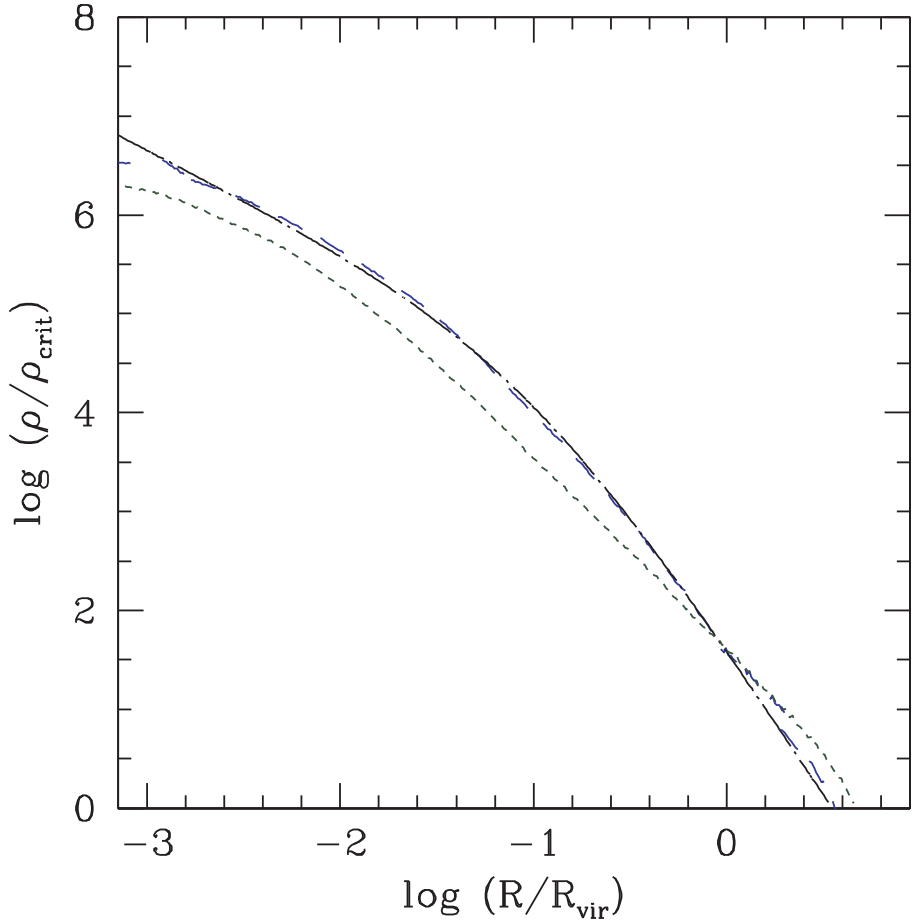,width=12.0cm}
\caption{Dark matter haloes generated with the model of Section 2. The dotted-dashed line is the NFW profile, the short-dashed line is the profile for the $10^{12} h^{-1} M_{\odot}$ halo obtained by means of our model. The dashed line represents the density profile obtained from the $10^{12} h^{-1} M_{\odot}$ halo
by reducing the magnitude of $h$, $j$ and $\mu$ as described in the text.
}
\end{figure}

%
%

\begin{figure}[ht]
\psfig{file=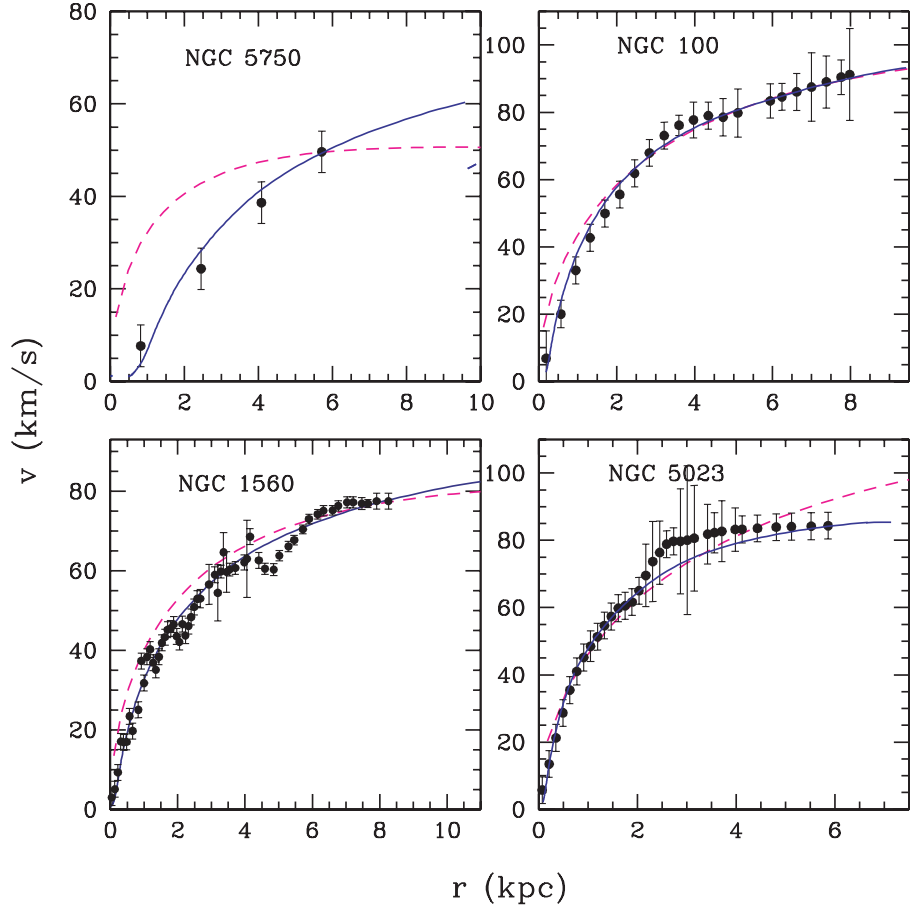,width=12.0cm}
\caption[]{Comparison of the rotation curves obtained with the model in Section 2 (solid lines) with the rotation curves of four LSB galaxies  studied by de Blok \& Bosma (2002).
The dashed line represents the fit with NFW model (see Section 3 for details). 
}
\end{figure}

\begin{figure}[tbp]
\centerline{\hbox{(a)
\psfig{figure=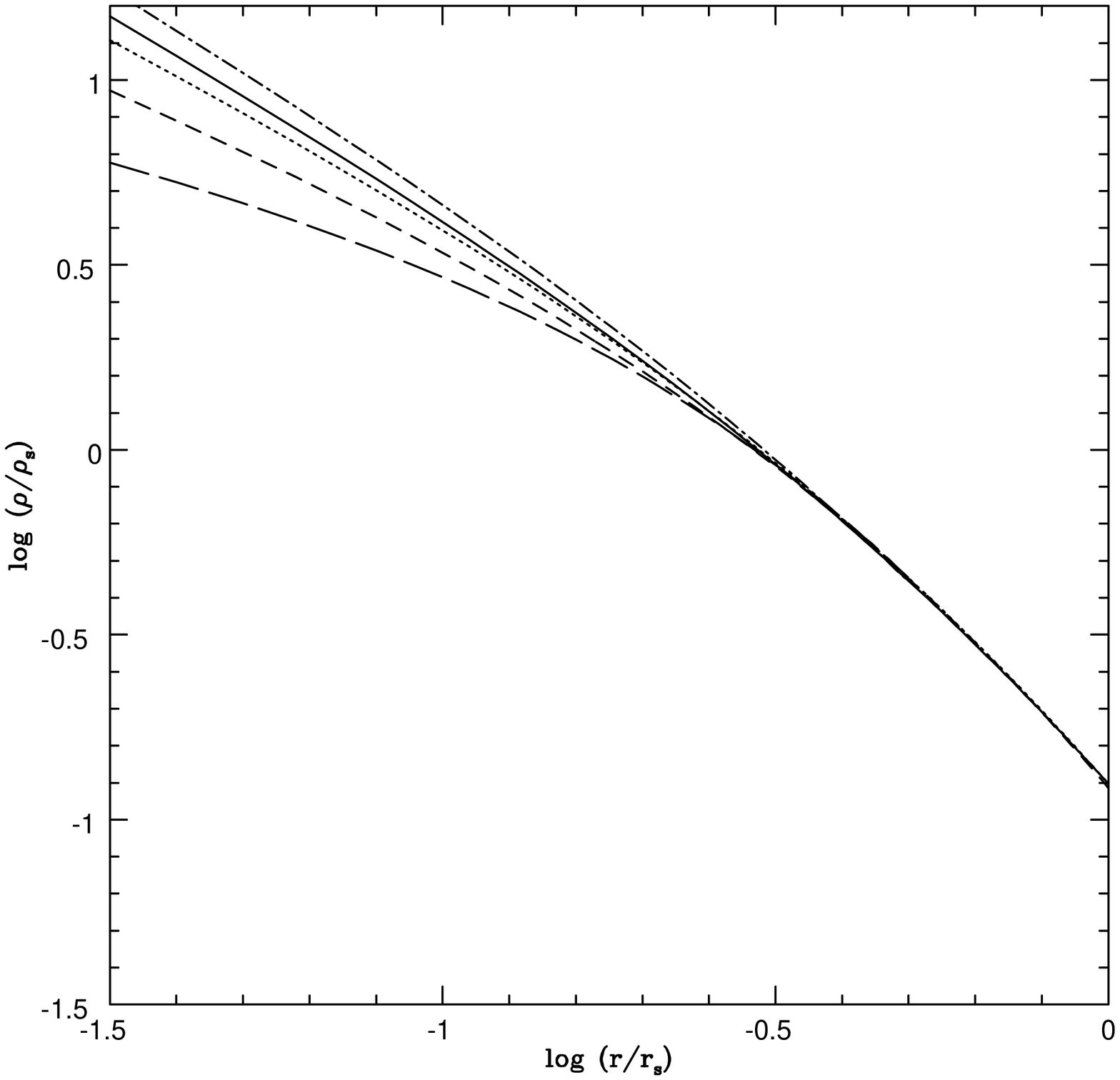,width=10cm} 
}}
\centerline{\hbox{(b)
\psfig{figure=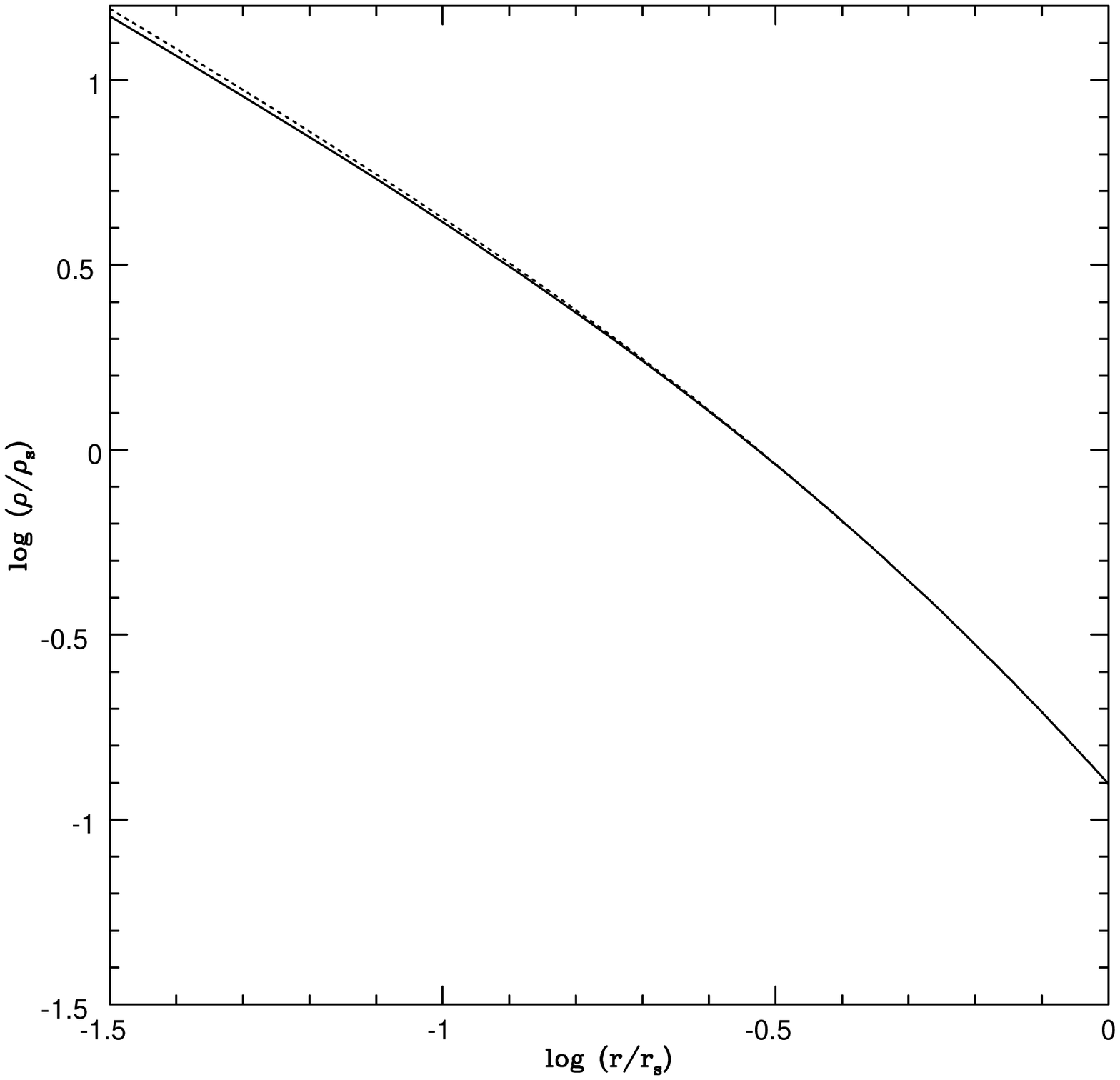,width=10cm}
}}
\caption{Panel (a) Density profile evolution of a $10^{14} h^{-1} M_{\odot}$ halo.
The solid line represents the NFW initial profile at $z=3$. Notice how the baryonic infall at redshift $2<z<3$ steepens the NFW cusp. 
The profile at $z=2$, $z=1.5$, $z=1$ and $z=0$ is represented by the dot-short-dashed line, dotted line, short-dashed line, 
and long-dashed line respectively.
Panel (b). Density profile evolution of a $10^{14} h^{-1} M_{\odot}$ halo for the total mass.
The solid line represents the NFW initial profile at $z=3$. 
The profile at $z=0$ is represented by the dotted line. 
}
\end{figure}

%

%

For what concerns the universal baryon fraction, it has been inferred by observations involving different physical processes (Turner 2002). The power spectrum of matter inhomogeneities of large-scale structure is sensitive to the quoted $F$; the Two Degree Field Galaxy Redshift Survey has reported a value of $0.15 \pm 0.07$ (Percival et al. 2001). Measurements of the angular power spectrum of the CMBR also provide a very significant estimate. The combined analysis in Jaffe et al. (2001) of several data gives $F= 0.186^{+0.010}_{-0.008}$. By using Wmap collaboration data (Spergel et al. 2003) one obtains $F \simeq 0.16$ ($\Omega_b=0.047 \pm 0.006$; $\Omega_m=0.29 \pm 0.07$). After baryons cools and form stars, the baryon-to-total mass at the virial radius 
does not need to be equal to the previous quoted universal baryon fraction and may deviate from it depending on feedback effects, hierarchical formation details, and heating by the extragalactic UVB flux.
In the AC calculation one has to take account that for large galaxies, only about half of the baryons, in principle available within the virial radius, goes into the central galaxy. For dwarf galaxies, this amount may be much smaller due to feedback effects, hierarchical formation details, and heating by the extragalactic UVB flux.
In the following, in the case of large galaxies, we use one half of the universal baryonic fraction and for dwarfs values obtainable from Hoeft et al. (2007) (their Fig. 1). For example in the case of a dwarf galaxy of $10^{10}M_{\odot}$ it is about 0.06. 

\section{Results and discussion}

After fixing the initial conditions and describing how to calculate angular momentum, dynamical friction and adiabatic contraction, we can use 
the model in Section 2 to obtain the density profile of haloes. 
In Fig. 1, we plot the profiles obtained through our model and those predicted by numerical simulations.
In Fig. 1, the short-dashed line, the dashed line, and the solid line represents, respectively, the density profile for haloes of 
$10^{12} h^{-1} M_{\odot}$, $10^{10} h^{-1} M_{\odot}$, and $10^{8} h^{-1} M_{\odot}$ calculated by means of the model of this paper. 
In order to compare the results of the model to those of N-body simulations, we plotted  
the NFW profile (dot-dashed line) for haloes having masses equal to $10^{12} h^{-1} M_{\odot}$, and $c=10$, and the results for the same mass 
obtained in the Aquarius Project by  Navarro et al. (2008)
(upper dashed-lines). Navarro et al. (2008) showed that density profiles deviate slightly but systematically from
the NFW model, and are better approximated by a fitting formula where the logarithmic slope is a power-law of radius: the Einasto profile, 
\begin{equation}
\ln(\rho(r)/\rho_{-2})=-2/\alpha [(r/r_{-2})^{\alpha}-1],
\end{equation}
where $r_{-2}$ and $\rho_{-2}$ are connected to the scaling radius and density of the NFW profile (see the following) by
$r_{-2}=r_s$ and $\rho_{-2}=\rho_s/4$.

Density profiles become monotonically shallower inwards, down to the innermost resolved point, with no indication
that they approach power-law behavior. The innermost slope they measure is slightly shallower than -1, a result
supported by estimates of the maximum possible asymptotic inner slope. 
Shallower cusps, such as the $r^{-0.75}$ behavior predicted by the model of Taylor
\& Navarro (2001), cannot yet be excluded.
The NFW profile for the given mass was calculated by means of the relationships connecting the concentration parameter, $c$, and the virial mass, $M_v$, to the shape of NFW profile. 
We used the following equation for $c$,
\begin{equation}
c \simeq 13.6 \left( \frac{M_v}{10^{11} M_{\odot}} \right)^{-0.13},
\label{eq:cvirr}
\end{equation} 
(Gentile et al. 2007),
and the usual one for the NFW profile,
\begin{equation}
\rho(r)=\frac{\rho_s}{r/r_s(1+r/r_s)^2}=\frac{\rho_b \delta_v}{r/r_s(1+r/r_s)^2}
\label{eq:navar}
\end{equation}
where
\begin{equation}
\delta_v=\frac{\Delta_v}{3}
\frac{c^3}{\log(1+c)-c/(1+c)},
\label{eq:navarr}
\end{equation}
and $\Delta_v$ is the virial overdensity (see Bryan \& Norman 1998).
The scaling radius, $r_s$, of the NFW profile is connected to the virial radius, concentration parameter and virial mass through $c=r_v/r_s$, where,
\begin{equation}
r_s \simeq 8.8 \left( \frac{M_v}{10^{11} M_{\odot}} \right)^{0.46}  {\rm kpc }
\end{equation} 
(Gentile et al. 2007).
The residuals from the best Einasto fits are extremely
small, and show no sign of systematic deviations down to
the innermost resolved point.
NFW residuals are then less than 30 \% over the full
radial range and below 20\% within $r_{-2}$. 

Figure 1 shows that halos generated using our method are different in character from the
profiles predicted by numerical simulations, like those of NFW or the Aquarius Project. 
This result is not surprising
since the types of evolution that numerical N-body and our halos undergo are rather different:
the former are produced as a cumulative result of many minor and major mergers of smaller
sub-halos, while the latter are the product of quiescent accretion of lumpy matter onto the
primary halo.
The NFW profiles change slope rapidly from $\alpha=1$ to $\alpha=3$ at the characteristic radius $r_s$. 
For example, in the case of the $10^{12} h^{-1} M_{\odot}$, the NFW profile have a characteristic scale length, equal to $0.1 r_v$, 
beyond which the density profile steepens, so that much of the mass is piled up within 10\% of the virial radius.
It is interesting to note that Navarro et al. (2004), Stadel et al. (2008) and the Aquarius Experiment start to show that 
the spherically-averaged density profile is not described, as in the NFW model just by an outer power-law $\rho \propto r^{-3}$ and inner power-law 
$\rho \propto r^{-1}$  but becomes progressively shallower inwards, and at the innermost resolved radius, the logarithmic slope is 
$\gamma \equiv -d \ln \rho /d \ln r \preceq 1$, ruling out claims of a steep $\rho \propto r^{-1.2}$ central cusp, and not excluding shallower cusps as the 
$r^{-0.75}$ behavior predicted by the model of Taylor and Navarro (2001). In other words, simulations with higher resolution are merging to the conclusion that the inner slope is not 
so steep as it appeared in less recent simulations and so in some cases even marginally consistent
the DM halo of NGC 4605 ($\rho \propto r^{-0.65}$) (Simon et al. 2003). In the following, we discuss another reason why density profiles obtained in N-body simulations are different from those obtained in our model.

NFW profiles change slope rapidly from $\alpha=1$ to $\alpha=3$ at the characteristic radius $r_s$. 
For example, in the case of the $10^{12} h^{-1} M_{\odot}$ model, the NFW profile has a characteristic scale length, equal to $0.1 r_v$, 
beyond which the density profile steepens, so that much of the mass is piled up within 10\% of the virial radius.
The haloes obtained using the model in Section 2 are different in character from the profiles predicted by numerical simulations, like those of NFW. 
Within the virial radius the log-log density slope changes gradually and the slopes of the inner part of haloes flatten with decreasing mass.
Our results show a steepening of the density profile with increasing mass with slopes $\alpha \simeq 0$ for $M \simeq 10^{8}- 10^{10} h^{-1} M_{\odot}$,
and $\alpha \simeq 0.8$ for $M \simeq 10^{12} h^{-1} M_{\odot}$.
The profiles with $M \simeq 10^{8}-10^{10} h^{-1} M_{\odot}$ are well fitted 
by means of a Burkert's profile considered a good fit to the dark matter rotation curves inferred from observations (e.g., Salucci \& Burkert 2000).
The functional form of this profile is characterized by,
\begin{equation}
\rho(r)= \frac{\rho_o}{(1+r/r_o)[1+(r/r_o)^2]},
\end{equation}
where $\rho_o \simeq \rho_s$ and $r_o \simeq r_s$ (EZ01).
The dark matter within the core is given by $M_o=1.6 \rho_o r_o^3$.
Although the dark matter parameters $r_o$, $\rho_o$ and $M_o$ are in principle independent, the observations reveal a clear connection (Burkert 1995):
\begin{equation}
M_o= 4.3 \times 10^7 \left ( r_o/kpc\right)^{7/3} M_{\odot}.
\end{equation}

In panel (a) of Fig. 1, we plot the result of the model of this paper in the case baryonic infall is not taken into account, while in panel (b) baryonic infall is taken into account. Comparing panel (a) and (b), we observe that the presence of baryons leads to a steeper profile in agreement with previous studies (Blumenthal et al. 1986; Flores et al. 1993; Klypin et al. 2002; Gnedin et al. 2004; Gustafsson et al. 2006).   
It is important to note that the effect of baryonic infall is more evident at early redshifts, as shown in Fig. 5. As shown in Fig. 5, from $z=3$ to $z=2$ the profile becomes even steeper than the NFW cusp. Later the effect of angular momentum and dynamical friction reduces the slope and erases the cusp. However, as shown in Fig. 1 the initial steepening of the profile remains visible in the profiles calculated through the model of the present paper ($M \simeq 10^{12} h^{-1} M_{\odot}$, $M \simeq 10^{10} h^{-1} M_{\odot}$, and $M \simeq 10^{8} h^{-1} M_{\odot}$), since the profile which takes into account 
baryonic infall (panel b) are steeper than those not taking it into account (panel a). 
\\

The model of the present paper gives just one profile for a given halo mass. In reality N-body simulations and analytical models different 
from those of Ryden-Williams and that of this paper (e.g. the extended Press-Schechter formalism), predict an ensemble of profiles for a given halo mass, depending on the different formation histories. 
For example, 
using the extended Press-Schechter (EPS) conditional probabilities for halo progenitor masses one can construct detailed histories of the mass
assembly of dark matter haloes. 
One is interested in computing the "mass accretion histories" (hereafter MAHs) of dark matter haloes, defined as $\Psi(M_0,z)=M(z)/M_0$, where $M(z)$ is defined as the mass of the "main progenitor halo", and $M_0$ is the halo mass at $z = 0$.
%
%
One then defines the average mass accretion history 
of a halo of mass $M_0$ as,
\begin{equation}
\Psi(M_0,z)= 1/N \sum_{i=1}^N \Psi_i (M_0,z),
\end{equation}
where the summation is over an ensemble of N random realizations (see van den Bosch \& Frank 2002).
One then can convert the mass growth curves $M(z)$ to density profiles as for example shown by Nusser \& Sheth (1999).
The single profile that we obtain in the present paper must be considered as the average profile of simulations.



Two important things must be noticed: a) less massive haloes are less concentrated; b) the halo's inner slope is smaller for smaller mass. \\

The first point can be explained as follows: higher peaks (larger $\nu$), which are progenitors of more massive haloes, have greater density contrast at their center, and so shells do not expand far before beginning to collapse. This reduces $j$ and $h$ and allows haloes to become more concentrated. An alternative explanation is connected to the quoted angular momentum-density anti-correlation showed
by Hoffman (1986): $j \propto \nu ^{-3/2}$ (and similarly for $h$). So, density peaks having low (high) value of $\nu $ acquire a larger (smaller) angular momentum than high 
$\nu $ peaks and consequently the halo will be less (more) concentrated. It is important to notice that the quoted trend of increased central concentration as a function of mass applies only to halos that started out as peaks in the density field smoothed with a fixed $R_f$ scale. Our conclusions do not mean that, for example, clusters of galaxies will be very much more centrally concentrated 
than galaxies, since different smoothing scales would apply in the two cases.

Point (b) can be explained in a similar way to (a), as described previously. 
Less massive objects are generated by peaks with smaller $\nu$, which acquire 
more angular momentum ($h$ and $j$). The angular momentum sets the shape of the density profile at the inner regions. For pure radial orbits, the core is dominated by particles from the outer shells. As the angular momentum increases, these particles remains closer to the maximum radius, 
resulting in a shallower density profile. 
Particles with smaller angular momentum will be able to enter the core but with a reduced radial velocity compared with the purely 
radial SIM. For some particles the angular momentum is so large that they will never fall into the core (their rotational kinetic energy makes them unbound). Summarizing, particles with larger angular momenta are prevented from coming close to the halo's center and so contributing to the central density. This has the effect of flattening the density profile.
This result is in agreement with  
the previrialization conjecture (Peebles \& Groth 1976; Davis
\& Peebles 1977; Peebles 1990), according to which initial asphericities and tidal
interactions between neighboring density fluctuations induce
significant non-radial motions that oppose the collapse. 
%
%
In order to reproduce the NFW profile, we performed an experiment similar to that performed by Williams et al. (2004), namely we 
reduced the magnitude of the $h$ and $j$ angular momentum, dynamical friction, $\mu$ and we considered the system as 
constituted only of dark matter ($F=0$). 
The experiment was performed on the halo of mass $10^{12} h^{-1} M_{\odot}$, and in order to reproduce the NFW profile having $c=10$ and mass 
$\simeq 10^{12} h^{-1} M_{\odot}$,
we had to reduce the magnitude of $h$ by a factor of 2, $j$ and $\mu$ by a factor of 2.5. 
The result of the quoted experiment is the dashed line in Fig. 2, which closely reproduces the NFW profile (dot-dashed line), 
and the dotted line is the profile for the $10^{12} h^{-1} M_{\odot}$ halo obtained by means of our model.\footnote{Note that the density profile 
of the Aquarius Experiment can be reproduced in the same  manner with a small change in the $h$ and $j$ parameters used for the NFW profile. 
}.

Similarly, Williams et al. (2004) had to 
reduce random velocities, which amount to reducing the angular momentum, in order to obtain a NFW profile. 
With each reduction of the random velocities, the profiles get steeper at the center. This effect can be understood, as already reported, as follows: the central density 
is built up by shells whose pericenters are very close to the center of the halo. Particles with larger angular momenta are prevented from 
coming close to the halo's center and so contributing to the central density. 
The correlation between increasing angular momentum and the reduction of inner slopes in halos has been also noticed by several other authors 
(Avila-Reese et al. 1998, 2001; Subramanian et al. 1999; Nusser 2001; Hiotelis 2002; Le Delliou \& Henriksen 2003; Ascasibar et al. 2004).  
As previously mentioned, there could be, according to Williams et al. (2004) another reason of the difference of density profiles obtained from N-body simulations and models like that of this paper (or William's).
As shown in their Fig. 1 (lower-left panel), the specific angular momentum distribution (SAM) in 
NFW-like halo is more centrally concentrated than the SAM distribution of the reference halo obtained with their model, model which 
is similar to that of this paper,
and is closer to those of typical halos emerging from numerical simulations.
This may suggest
that haloes in N-body simulations lose a considerable amount of 
angular momentum between 0.1 and 1 $r_{v}$. 
Since virialization proceeds from inside out, this means that the angular momentum loss takes 
place during the later stages of the halos' evolution, rather then very early on. 
This is somehow confirmed by the so called angular momentum catastrophe, 
namely the fact that dark matter halos generated through
gas-dynamical simulations
are too small and have too little angular momentum compared to the halos of real disk galaxies, possibly because it was lost during repeated collisions through dynamical friction or other mechanisms (van den Bosch et al. 2002; Navarro \& Steinmetz 2000). 
The problem can be solved invoking stellar feedback processes (Weil et al. 1998), but part of the angular momentum problem seems due to numerical effects, most likely related to the shock capturing, artificial viscosity used in smoothed particle hydrodynamics (SPH) simulations (Sommer-Larsen \&Dolgov 2001).

We discussed the effect of changing the magnitude of angular momentum but we did not speak of the effect of changing the magnitude 
of $\mu$ (dynamical friction). The effect of changing this last quantity is very similar to changing the magnitude of angular momentum: 
an increase in the term $\mu$ produces shallower profiles as larger values of angular momentum does. This is expected from Del Popolo (2006) (Fig. 1), showing that 
dynamical friction influence the dynamics of collapse in a similar way to that
of angular momentum slowing down the collapse of outer shells and so compelling the particles to remain closer to the maximum radius. 

At this point, we have to stress that
the flattening of the density profile is caused by two main effects: angular momentum and dynamical friction. The other effect that we have considered, namely baryonic infall, counteracts the effect of the density profile flattening produced by angular momentum and dynamical friction. The role of baryonic infall is predominant at early times and produces a steepening of the profile, while later its effect is overwhelmed by the effect of angular momentum and dynamical friction. 
The result is similar to the one described by Williams et al. (2004) with the difference that in the present paper the effect of ordered angular momentum and dynamical friction adds to that of random angular momentum studied by Williams et al. (2004) with the result that the flattening of the density profiles is larger. 

In Fig. 3, we plot the rotation curves obtained by our model and we compare them to four LSB galaxies shown in Williams et al. (2004).
The four LSBs in Fig. 3 were taken from the high-resolution data of de Blok \& Bosma (2002). Of the 26 LSBs that are
presented in that paper we picked the ones with high inclination angles, smooth rotation
curves, and good agreement between HI, Ha, and optical data.
In all the four cases, the data are compared with 
the rotation curve obtained using our model (solid line) and with rotation curves obtained from NFW profile (dotted lines), given by,  
\begin{equation}
V(r)=V_v
\left \{
\frac{\ln (1+cx)-cx/(1+cx)}
{x [\ln(1+c)-c/(1+c)]}
\right \}^{1/2}.
\end{equation}
where $x=r/r_v$ and $V_v$ is the virial velocity \footnote{The value of the characteristic velocity $V_v$ of the halo is defined in the same way as the virial radius $r_v$. 
}. 
The concentration parameter $c$ was chosen to be consistent with the NFW predictions, calculating it from the mass of the galaxy through 
Eq. (\ref{eq:cvirr}).
For all the four galaxies the rotation curves obtained with the model of the present paper (solid lines) are a considerably better fit than NFW (dotted lines).
Fig. 3 shows that NFW haloes are higher than the rotation curves obtained using our model, in which more massive haloes tend to be more centrally concentrated and have flatter rotation curves. Less massive haloes are less concentrated, and have slowly rising rotation curves. In contrast
NFW rotation curves rise very steeply and as a consequence NFW fits to dwarf galaxy rotation curves have too low concentration parameters (van den Bosch \& Swaters 2001) compared to N-body predictions. 
NFW fails to reproduce the velocities and the shape of the observed rotation curves,
predicting too high velocities in the central part of haloes, and  
even leaving $c$ as free parameter, instead of using Eq. (\ref{eq:cvirr}), there is no appreciable improvement in the fit. Using Eq. (\ref{eq:cvirr}), one obtains very low values of $c$. 
The result is similar to that described by Gentile et al. (2004): data are much better described by core-like profiles, like the Burkert profile generating flatter rotation curves,
\begin{equation}
V(r)= \left 
(\frac{2 \pi G \rho_0 r_o^3}{r})
\right)^{1/2}
\left \{
\ln
\left[
(1+r/r_o) \sqrt(1+(r/r_o)^2)
\right]-
\arctan (r/r_o)
\right \}^{1/2},
\end{equation}
Our rotation curves are very similar to those generated by the Burkert profile and the residuals and discrepant points for our rotation curves are close to that given in Gentile et al. (2004) for the Burkert's fit to their data.

Our halos appear to be a closer match to the halos of spiral and dwarf
galaxies, than are N-body halos. This may indicate that the halos of real late-type disk
galaxies undergo a formation scenario similar to the one depicted by our method, i.e. collapse
proceeds through a quiescent accretion of lumpy material and minor mergers, rather than
through a merger-driven formation process characteristic of fully hierarchical models.

%
Before going on, we notice that of the four rotation curves, NGC100 
seems to be almost equally well fitted by the two type of models. 
In some other cases, the difference between the NFW rotation curves and those of the present paper is even smaller than for NGC100 and so it would be worthwhile to discuss halo-to-halo scatter. During the hierarchical assembly of dark matter haloes, the inner regions of early virialized objects often survive accretion onto a larger system, thus giving rise to a population
of subhaloes. Depending on their orbits and their masses, these subhaloes therefore either merge,
are disrupted or survive to the present day. To fully describe, in a statistical sense, the non-linear
distribution of mass in the Universe, it is essential that halo substructure is taken into account. 
These subhalos appear ubiquitous in high resolution cosmological simulations and provide the source of fluctuations.
One can study the evolution of halos under the influence of a
generation of subhalos, while real halos grow
continuously by accretion and mergers. 
Fluctuations due to subhalos
in parent halos are important for understanding the
time evolution of dark matter density profiles and the
halo-to-halo scatter of the inner cusp seen in recent ultrahigh
resolution cosmological simulations. 
This scatter may be explained by subhalo
accretion histories: when we allow for a population of
subhalos of varying concentration and mass, the total
inner profile of dark matter can either steepen or flatten.

As noticed in the introduction, while on galactic scales a large number of studies predicts central cores and it seems that the cusp/core problem
is a real problem not attributable to systematic errors in the data (de Blok, Bosma \& McGaugh 2003), on cluster scales the situation is less clear with slopes ranging from $\alpha=0.5$ (e.g. Sand et al. 2002; Sand et al. 2004) to $\alpha=1.9$ (Arabadjis et al. 2002).


In order to study the problem on cluster scales, we have calculated the density profile evolution of dark matter and that of the total matter 
distribution for a halo of $\simeq 10^{14} h^{-1} M_{\odot}$.

We used the method of Sect. 2 to calculate the density profile and then we repeated the experiment shown in Sect. 3 to reproduce the NFW profile for the quoted mass that we considered as the initial condition. \footnote{We could have directly started from a NFW profile for the given mass calculated by means of the NFW fitting formula.} 

After we evolved this profile using the method of Sect. 2 and taking the redshift dependence in the model using the technique described in Del Popolo (2001) (the reader is referred to the quoted paper to have more insights). 
The goal was to study the evolution of a NFW profile similarly to El-Zant et al. (2001,2004), Tonini et al. (2006), Romano-Diaz et al. (2008). 


Before showing the results, we recall that in real haloes angular momentum is lost in the final stages of collapse through the transfer of angular momentum from the subclumps to external material through dynamical friction (Quinn \& Zurek 1988; R88b; Klypin et al. 2002). 
Once the baryons condense to form stars and galaxies, they experience a dynamical friction force from the less massive dark matter particles as they move through the halo. Energy and angular momentum is transferred to dark matter, increasing its random motion. Moreover, angular momentum acquired 
in the expansion phase gives rise to non-radial motions in the collapse phase. 
The spherical approximation that we use, which ignores the possibility of large substructure forming, neglects the transfer of angular momentum through the interactions of subclumps 
\footnote{An approximate analytical approach to the problem of exchange of angular momentum between the dark matter particles and infalling baryons has been developed by 
Klypin et al. (2002).}.
In the present paper, the effect of dynamical friction was just taken into account as an additive term in the equation of motion, Eq. (\ref{eq:coll}),
without changing the sphericall symmetry of the problem as done by Peebles (1993) and Del Popolo (2006).
The effect of the dynamical friction term in the equation of motion is very similar to that of angular momentum: 
an increase in the term $\mu$ produces shallower profiles as larger values of angular momentum does. This is expected from Fig. 1 of Del Popolo (2006), showing that dynamical friction influences the dynamics of collapse in a similar way to that of angular momentum slowing down the collapse of outer shells and so compelling the particles to remain closer to the maximum radius. 

Fig. 4a plots the evolution of a density profile of $10^{14} h^{-1} M_{\odot}$. The solid line represents the initial density profile at $z=3$, which slightly steepens due to baryon settling in virialized dark matter haloes (AC) (dot-short-dashed line) at $z=2$. 
Moreover, angular momentum acquired 
in the expansion phase gives rise to non-radial motions in the collapse phase. The effect of angular momenta and dynamical friction overcomes that of the AC and the profile starts to flatten ($z=1.5$ dotted line; $z=1$ short-dashed line; $z=0$ long-dashed line). The final dark matter profile (long-dashed line) is characterized by a log-log slope of $\alpha \simeq 0.6$ at $0.01 r_s$. So the situation is similar to that of haloes on galactic scales but the slope is larger than for dwarf galaxies. 
In Fig. 4b, we plot the evolution of the total density profile at initial redshift, $z=3$, (solid line) and at the final redshift $z=0$ (dotted line). The plot shows that the total density profile slightly changes 
with time and the cusp is not ``erased" as in the case of the dark matter profile. This result also implies that the baryonic component becomes steeper than the original NFW profile. The behavior of total mass is in agreement with X-ray observations by Chandra and XMM (Buote 2003, 2004; Lewis et al. 2003), weak lensing (Dahle et al. 2003) and strong lensing (Bartelmann 2003), which are consistent with a cusp having $\alpha=1$ or larger. The behavior of the dark matter halo is in agreement with analysis of Sand et al. (2002, 2004) 
who fitted the baryonic and dark matter profiles only in the very inner part of the cluster MS 2137-23 within $\simeq 50 h^{-1} $ kpc by means of a generalized NFW profile:
\begin{equation}
\rho(r)= \frac{\rho_b \delta_v}{(r/r_s)^{\alpha}(1+r/rs)^{3-\alpha}}
\end{equation} 
obtaining a nearly flat core $\alpha=0.35$. The steepening of the baryonic component is consistent with what found by Brunzendorf \& Meusinger (1999), who found that the projected galaxy distribution in the Perseus cluster diverges as $r^{-1}$.


The results previously reported have several implications on the effort to test predictions of the CDM model observationally. 
The test that received much attention in the last decade, as several times stressed, is the density distribution in the inner regions of galaxies and clusters. 
On galactic scales, as our results show, the infall of baryons at early times steepen the cusp due to AC (in agreement with previous studies: Blumenthal et al. 1986; Gnedin et al. 2004; Gustafsson et al. 2006) but later the cusp is erased through dynamical friction and non-radial motions effects. This results are in agreement with other analyses which study separately the effects of dynamical friction (e.g., EZ01) and those of angular momentum (non-radial motions) (e.g., Nusser 2001; Hiotelis 2002; Ascasibar et al. 2004; Williams et al. 2004) and with the recent Sph simulations of Romano-Diaz et al. (2008).
%
%

Going to larger scales the situation changes. The analysis of the density distribution for the bright galaxies is complicated by the uncertain contribution of stars to the total mass profile (Treu \& Koopmans 2002; Mamon \& Lokas 2004). Some analyses tend to favor inner slopes shallower than predicted by CDM (e.g., Gentile et al. 2004) but others deduce slopes of the inner profiles that are at least marginally consistent with predictions (Treu \& Koopmans 2002, 204; Koopmans \& Treu 2003; Jimenez et al. 2003). 
As previously reported, our results shows a steepening of the density profile with increasing mass with a density profile of haloes of mass $ > 10^{12} h^{-1} M_{\odot}$ having slopes $>0.8$. 
This is in agreement with recent N-body simulations having a 
logarithmic slope that decreases inward more gradually than the NFW profile (Hayashi et al. 2003; Navarro et al. 2004; Stadel et al. 2008).  
In the case of Stadel et al. (2008) the logarithmic slope is $0.8$ at $0.05 \%$ of $r_v$.

The density distribution in clusters of galaxies can, in principle, provide a cleaner test of the models because the effects 
of the baryons and gas on the dark matter distribution are expected to be smaller and simpler. However also in this case observations predict slopes ranging from $\alpha \leq 0.5$ ($\alpha=0.35$,  Sand et al. 2002, 2004) to values larger than one (Arabadjis et al. 2002) with in some cases different results even for the same object. This is the case of the cluster MS2137-23 studied by Sand et al. (2004), who found a shallow density slope ($\leq 0.5$)  
while Dalal \& Keeton (2003), Bartelman \& Meneghetti (2004) and Gavazzi (2003) contested Sand's results, which according to them is neglecting lens ellipticity, and found consistency of the inner slope with a NFW profile. 
Our result concerning cluster scales makes a difference between dark matter and total mass distribution: the first tends to be less cuspy than the NFW profile in agreement with some observations (e.g., Sand et al. 2002, 2004), while the second is a bit more cuspy than the 
NFW profile (in agreement with Brunzendorf \& Meusinger 1999) . 


In the present paper, we did not take into account the baryon-DM interaction. As previously reported
once the baryons condense to form stars and galaxies, 
energy and angular momentum is transferred to dark matter, increasing its random motion. 

In future work, 
we will develop a semianalytic
model to follow the evolution of the baryonic component, and its 
interaction with DM, and we will compare these with full hydrodynamical
simulations.
This should bring new insights into galaxy formation, and directly address possible small-scale
challenges to the CDM theory. 
The role of the interaction between DM and baryons could bring further changes in the final density profile, that we shortly discuss. 
It is interesting to note that the problems of CDM only become clear on length scales where the
baryons start playing a role and that this applies not only to the cusp/core problem but
also to the missing dwarfs problem. Moreover, it is noteworthy that if, indeed, most 
starforming galaxies in the early universe lost their DM cusps because of stellar feedback, the
missing dwarfs (satellites) problem could also be solved. Dwarf galaxies without a central
cusp have a lower average core density than cuspy ones, and are hence much easier to disrupt
tidally during the hierarchical assembly of larger galaxies (Mashchenko \& Sills 2005). As a
consequence, the removal of galactic cusps by stellar feedback in the early universe would
result in fewer satellites today. This again indicates that baryon physics could be one of the missing
pieces of the puzzle, and will very likely make a major contribution toward a solution. If this
is true, it would be unwise to ignore the conclusions to which data are leading us, namely
that small scales tells us more about galaxy formation than it does about 
CDM\footnote{Other possibilities are that the observed dark matter cores are telling us that dark matter has pressure
at small scales or something unexpected.}. In other
terms, the centers of galaxies are special places, the only places where we can study dark
matter under peculiar conditions.

%


\section{Conclusions}

In this paper, we studied the cusp/core problem by improving the ESIM of Williams et al. (2004).
We took into account, simultaneously, the effects of ordered and random angular momentum, dynamical friction and adiabatic contraction. 
The improved SIM of the present paper, taking account the previous effects gives rise to haloes being characterized by a log-log density slope that changes gradually within the virial radius and slopes of the inner part of haloes flattening with decreasing mass. As in 
previous papers, AC steepens the density profiles.
The density profiles of structure having masses smaller than $10^{11} h^{-1} M_{\odot}$ are well fitted by Burkert's profiles. The comparison of some of the rotational curves given by de Blok \& Bosma (2002)
with the rotational curves obtained by means of our model, gives a good agreement. 
In the case of clusters of galaxies the density profile evolution is similar to that observed on galactic scales with the difference that the final slope is steeper than in the dwarf galaxies case. However the total mass profile is still cuspy and evolves slightly.
The behavior of the dark matter halo is in agreement with the analysis of Sand et al. (2002, 2004) who fitted the baryonic and dark matter profiles only in the very inner part of the cluster MS 2137-23 within $\simeq 50 h^{-1} $ kpc by means of a generalized NFW profile. 
The behavior of total mass is in agreement with X-ray observations by Chandra and XMM (Buote 2003, 2004; Lewis et al. 2003) weak lensing (Dahle et al. 2003) and strong lensing (Bartelmann 2003), which are consistent with a cusp having $\alpha=1$ or larger.
As also reported by Williams et al. (2004), numerically generated halos must have lost a considerable amount of their angular momentum in the outer parts, roughly between 0.1 and $1 R_v$, possibly through dynamical friction or other mechanisms.
A further future model taking into account the interaction between DM and baryons would help to better understand the problem.


 
\acknowledgements

We would like to thank Massimo Ricotti, Nicos Hiotelis, and Antonaldo Diaferio for their very helpful suggestions and comments.






\begin{thebibliography}{}

\bibitem{}Antonuccio-Delogu V., Colafrancesco S., 1994, ApJ 427, 72 (ADC)
\bibitem{} Arabadjis J. S., Bautz M. W., Garmire G. P., 2002, ApJ,
572, 66
\bibitem{} Ascasibar Y., Yepes G., Gottlober S., 2004, MNRAS 352, 1109A
\bibitem{} Ascasibar Y.,  Hoffman Y., Gottl\"ober S., 2007, MNRAS 376, 393
\bibitem{} Avila-Reese V., Firmani C., Hernandez X., 1998, ApJ, 505,
37
\bibitem{} Bardeen J.M., Bond J.R., Kaiser N., Szalay A.S., 1986, ApJ 304, 15
\bibitem{} Barnes, J., Efstathiou, G., 1987, ApJ 319, 575

\bibitem[]{} Bartelmann, M. 2003, in ``Matter and Energy in Clusters of Galaxies'', ASPC 301, 255 
\bibitem{} Bartelman M., Meneghetti M., 2004, A\&A 418, 413
\bibitem{} Bertschinger E., 1985, ApJS 58, 39
\bibitem{} Blumenthal G. R., Faber S. M., Flores R., Primack J. R.,
1986, ApJ, 301, 27
\bibitem{} Boissier S., Monnier R. D., van Driel W., Balkowski C., Prantzos N., 2003, Ap\&SS 284, 913
\bibitem{} Borriello A., Salucci P., 2001, MNRAS, 323, 285
\bibitem{} Bothun, G., Impey, C. \& McGaugh, S. 1997, PASP, 109, 745  
\bibitem{} Bryan G. L., Norman M. L., 1998, ApJ 495, 80
\bibitem{} Bullock, J.S., Dekel, A., Kolatt, T. S., Kravtsov, A.V., Klypin, A. A., 
Porciani, C. \& Primack, J. R. 2001, ApJ, 555, 240 
\bibitem{} Brunzendorf, J.  \& Meusinger, H.  1999,  Astron. \&  Astrophys. Suppl. Ser. , 139, 41
\bibitem[]{} Buote, D. A. 2003, in ``IGM/Galaxy Connection--The Distribution of Baryons at z=0'', University of Colorado, 2003,
ASSL Conference Proceedings Vol. 281, p. 87. 
\bibitem[]{} Buote, D. A. 2004, Proc. IAU 220, 149 
\bibitem{} Burkert A., 1995, ApJ, 447, L25
\bibitem{} Cardone V. F., Sereno M., 2005, A\&A 438, 545
\bibitem{}Catelan P., Theuns T., 1996, MNRAS2 82, 436
\bibitem{} Cen R., 2001, ApJ, 546, L77
\bibitem{} Cen R. Y., Dong F., Bode P., Ostriker J. P., 2004, astro-ph/0403352
\bibitem{} Cole S., Lacey C., 1996, MNRAS 281, 716
\bibitem{} Colin P., Avila-Reese V., Valenzuela O., 2000, ApJ 542,
622
\bibitem{} Dahle H., Hannestad S., Sommer-Larsen J., 2003, ApJ, 588,
L73
\bibitem{} Dalal N., Keeton C. R., 2003, astro-ph/0312072
\bibitem{} Davis M., Peebles P.J.E., 1977, ApJS, 34, 425
\bibitem{} Dav\'e R., Spergel D. N., Steinhardt P. J., Wandelt B. D.,
2001, ApJ, 547, 574
\bibitem{} de Blok, W.~J.~G., McGaugh, S.~S., Bosma, A., \& Rubin,
V.~C. 2001a, ApJ, 552, L23
\bibitem{} de Blok, W.~J.~G., McGaugh, S.~S., \& Rubin, V.~C. 2001b,
AJ, 122, 2396
\bibitem{} de Blok, W.~J.~G., Bosma, A., \& McGaugh, S.\ 2003, MNRAS
340, 657
\bibitem{} de Blok W. J. G., McGaugh S. S., Bosma A., Rubin V. C.,
2001, ApJ 552, L23
\bibitem{} de Blok W. J. G., Bosma A., 2002, A\&A 385, 816
\bibitem{} de Blok W. J. G.,  2003, in dark matter in Galaxies, ASP Conference series, Vol. 220, 2003, S. Ryder, D. J. Pisano, M. Walker, and K. C. Freeman, eds. 
\bibitem{}  Del Popolo A., Gambera M., Recami E., Spedicato E., 2000, A\&A 353, 427 (DP2000)
\bibitem{} Del Popolo A., 2006, A\&A 454, 17
\bibitem{} Dubinski J., Carlberg R., 1991, ApJ 378, 496
\bibitem{} Eisenstein D.J., Loeb A., 1995, ApJ 439, 250
\bibitem{} El-Zant A., Shlosman I., Hoffman Y., 2001, ApJ, 560, 636 (EZ01)
\bibitem{} El-Zant A., Hoffman Y., Primack J., Combes F., Shlosman
I., 2003, ApJ, 2004, ApJ 607, 75
\bibitem{} Ettori S., Fabian A. C., Allen S. W., Johnstone R. M.,
2002, MNRAS, 331, 635 

\bibitem{} Filmore J.A., Goldreich P., 1984, ApJ 281, 1
\bibitem{} Flores R. A., Primack J. R., 1994, ApJ, 427, L1
\bibitem{} Flores R., Primack J.~R., Blumenthal, G.~R., \& Faber, S.~M. 1993, ApJ 412, 443

\bibitem{} Fukushige T. \& Makino, J, 2001, ApJ 557, 533
\bibitem{} Fukushige T., Kawai A., Makino J. 2004, ApJ 606, 625 
\bibitem{} Ghigna S., Moore B., Governato F., Lake G., Quinn T., Stadel J., 2000, ApJ, 544, 616 
\bibitem{} Gavazzi R., 2003, in Impact of Gravitational Lensing on Cosmology Proceedings IAU Symposium N. 225, Mellier Y., \& Meylan G. eds.
\bibitem{} Gelato S., Sommer-Larsen J., 1999, MNRAS 303, 321
\bibitem{} Gentile G., Salucci P., Klein U., Vergani D., Kalberla P., 2004, MNRAS 351, 903
\bibitem{} Gentile G., Tonini C. \& Salucci P., 2007, MNRAS 378, 41
\bibitem{} Gnedin, O. Y., Kravtsov A. V., Klypin A. A., Nagai D., 2004, ApJ 616, 16
\bibitem{} Goodman J., 2000, New Astronomy 5, 103
\bibitem{} Gott J.R., 1975, ApJ 201, 296 
\bibitem{} Gunn J.E., Gott J.R., 1972, ApJ 176, 1
\bibitem{} Gunn J.E., 1977, ApJ 218, 592
\bibitem{} Hayashi E., Navarro J. F., Power C., Jenkins A., Frenk C. S., White S. D. M., Springel V., Stadel J., Quinn T. R., 2004 MNRAS, 355, 794
\bibitem{} Henriksen R. N., Widrow L. M., 1999, MNRAS 302, 321
\bibitem{} Hiotelis, N. 2002, A\&A 383, 84
\bibitem{} Hoeft M., Mucket J. P., Gottlober S., 2004, ApJ 602, 162
\bibitem{} Hoffman, Y., 1986, ApJ 301, 65
\bibitem{} Hoffman Y., Shaham J., 1985, ApJ 297, 16 (HS)
\bibitem{} Hoyle, F.: (1949), in IAU and International Union of Theorethicaland Applied Mechanics Symposium, p. 195
\bibitem{} Hozumi S., Burkert A., Fujiwara T., 2000, MNRAS 311, 377
\bibitem{} Hu W., Barkana R., Gruzinov A., 2000, Physical Review
Letters, 85, 1158
\bibitem{} Jimenez R., Verde L., Oh S. P., 2003, MNRAS 339, 243 
\bibitem{} Jing, Y. P. \& Suto, Y. 2000, ApJ, 529, L69  
\bibitem{} Kandrup, H.E., 1980, Phys. Rep. 63, n 1, 1
\bibitem{} Kaplinghat M., Knox L., Turner M. S., 2000, Physical Review Letters, 85, 3335
\bibitem{} Kashlinsky A., 1986, ApJ, 306, 374
\bibitem{} Kashlinsky A., 1987, ApJ, 312, 497
\bibitem{} Keeton C. R., 2001, ApJ 561, 46 
\bibitem{} Klypin, A., Kravtsov, A. V., Bullock, James S. \& Primack, J. R. 2001, ApJ, 554, 903 
\bibitem{} Klypin A., Zhao H., \& Somerville R.~S. 2002, ApJ 573, 597
\bibitem{} Koopmans L. V. E., Treu T., 2003, ApJ 583, 606
\bibitem{} Kravtsov A. V., Klypin A. A., Bullock J. S., Primack J. R.,
1998, ApJ, 502, 48
\bibitem{} Le Delliou M., Henriksen R. N., 2003, A\&A 408, 27
\bibitem{} Lemson, G., 1995, Ph.D. thesis, R\"{\i}ksuniversiteit Groningen  
\bibitem{} Lewis A. D., Buote D. A., Stocke J. T., 2003, ApJ 586,
135
\bibitem{} Mamon G.A., Lokas E. L., MNRAS 2005, 362, 95
\bibitem{} Marchesini D., D'Onghia E., Chincarini G., Firmani C.,
Conconi P., Molinari E., Zacchei A., 2002, ApJ, 575, 801
\bibitem{} Mashchenko S., Couchman H. M. P., Wadsley J., Nature 442, 539
\bibitem{} Mashchenko S. \& Sills,  2005, ApJ 619, 258.
\bibitem{} Merrit D., Navarro J. F., Ludlow A., Jenkins A., 2005, ApJ 624, L85
\bibitem{} Mo H.~J., Mao S., \& White S.~D.~M. 1998, MNRAS 295, 319
\bibitem{} Moore B., 1994, Nature, 370, 629
\bibitem{} Moore, B., Governato, F., Quinn, T., Stadel, J. Lake, G. 1998, ApJ, 499, L5 
\bibitem{} Navarro J.F., Frenk C.S., White S.D.M., 1995, MNRAS 275, 720
\bibitem{} Navarro J.F., Frenk C.S., White S.D.M., 1996, ApJ 462, 563
\bibitem{} Navarro J.F., Frenk C.S., White S.D.M., 1997, ApJ 490, 493 (NFW) 
\bibitem{} Navarro J. F., Hayashi E., Power C., Jenkins A. R., Frenk C. S., White S. D. M., Springel V., Stadel J., Quinn T. R, 2004,MNRAS 349, 1039
\bibitem{} Nusser A., Sheth R. K., 1999, MNRAS, 303, 685
\bibitem{} Nusser A., 2001, MNRAS 325, 1397
\bibitem{} Peebles, P. J. E., 1969, ApJ 155, 393
\bibitem{} Peebles P.J.E., Groth E.J., 1976, A\&A 53, 131
\bibitem{} Peebles P.J.E., 1980, The large scale structure of the Universe, Priceton University Press
\bibitem{} Peebles, P.J.E., 1990, ApJ 365, 27
\bibitem{} Peebles P. J. E., 2000, ApJ 534, L127
\bibitem{} Power, C., Navarro, J. F., Jenkins, A., Frenk, C. S., White, S. D. M., Springel V., Stadel J., Quinn T., 2003, MNRAS 338, 14
\bibitem{} Quinn P.J., Salmon J.K., Zurek W.H., 1986, Nature, 322, 329
\bibitem{} Rhee G., Valenzuela O., Klypin A., Holtzman J., Moorthy B., 2004, ApJ 617, 1059	
\bibitem{} Ricotti M., 2003, MNRAS 344, 1237
\bibitem{} Ricotti M., Wilkinson M. I., 2004, MNRAS 353, 867 
\bibitem{} Ricotti M., Pontzen A., Viel M., 2007, ApJ 663, 53
\bibitem{}  Rix, Hans-Walter; de Zeeuw, P. Tim; Cretton, Nicolas; van der Marel, Roeland P.; Carollo, C. Marcella,  1997, ApJ 488, 702 
\bibitem{} Romano-Diaz E., Shlosman I., Hoffman Y., Heller C., 2008, astro-ph 0808.0195
\bibitem{} Ryden B.S., Gunn J.E., 1987, ApJ 318, 15 (RG87)
\bibitem{} Ryden, B.S., 1988a, ApJ 329, 589 (R88)
\bibitem{} Ryden, B.S., 1988b, ApJ 333, 78
\bibitem{} Salucci P., Burkert A., 2000, ApJ 537, L9 
\bibitem{} Sand D. J., Treu T., Ellis R. S., 2002, ApJ 574, L129
\bibitem{} Sand D. J., Treu T., Smith G. P., Ellis R. S., 2004, ApJ 604, 88
\bibitem{} Simon J. D., Bolatto A. D., Leroy A. Blitz L., 2003b, in Satellites and Tidal Streams, ASP conference Series 2003
\bibitem{} Stadel J., Potter D., Moore B., Diemand J., Madau P., Zemp M., Kuhlen M., Quilis V, 2008, astro-ph/0808.2981
\bibitem{} Sommer-Larsen J., Dolgov A., 2001, ApJ 551, 608
\bibitem{} Spekkens K. Giovanelli R., Haynes M. P., 2005 AJ 129, 2119 
\bibitem{} Spergel D .N., et al. 2003, ApJS 148, 175
\bibitem{} Spergel D. N., Steinhardt P. J., 2000, Physical Review Letters 84, 3760
\bibitem{} Subramanian K., Cen R., Ostriker J. P., 2000, ApJ 538,
528
\bibitem{} Taylor J. E., Navarro J. F., 2001, ApJ 563, 483
\bibitem{} Taylor J. E., Silk J., Babul A., 2004, IAUS no. 220, p. 91
\bibitem{} Treu T. \& Koopmans L.~V.~E., 2002, ApJ 575, 87
\bibitem{} Treu T. \& Koopmans L.~V.~E., 2004, ApJ 611, 739
\bibitem{} van den Bosch, F.~C., \& Swaters, R.~A.\ 2001, \mnras 325,
1017
\bibitem{} van den Bosch, F.~C., Robertson, B.~E., Dalcanton, J.~J., \& de Blok, W.~J.~G.  2000, AJ 119, 1579
\bibitem{} Yoshida N., Springel V., White S. D. M., Tormen G., 2000, ApJ 544, L87
\bibitem{} White, S. D. M., 1984, ApJ 286, 38
\bibitem{} White S.D.M., Zaritsky D., 1992, ApJ 394, 1
\bibitem{} Williams L. L. R.,  Babul A.,  Dalcanton, J. J., 2004, ApJ 604, 18
\end{thebibliography}
\end{document}